\documentstyle[aps,prb,preprint,graphicx]{revtex}

\begin{document}
\tightenlines

\title{Development of an eight-band theory\\
for quantum-dot heterostructures}

\author{E. P. Pokatilov\cite{e-mail1} and V. A. Fonoberov\cite{e-mail2}}

\address{Laboratory of Multilayer Structure Physics, Department of
Theoretical Physics,\\ State University of Moldova, A. Mateevici
60, MD-2009 Chi{\c s}in{\u a}u, Moldova}

\author{V. M. Fomin\cite{perm.address} and J. T. Devreese\cite{d.address}}

\address{Theoretische Fysica van de Vaste Stof, Departement
Natuurkunde,\\ Universiteit Antwerpen (UIA), Universiteitsplein 1,
B-2610 Antwerpen, Belgi\"e}

\date{\today}

\maketitle

\begin{abstract}
We derive a nonsymmetrized 8-band effective-mass Hamiltonian for
quantum-dot heterostructures (QDHs) in Burt's envelope-function
representation. The 8$\times$8 radial Hamiltonian and the boundary
conditions for the Schr\"odinger equation are obtained for
spherical QDHs. Boundary conditions for symmetrized and
nonsymmetrized radial Hamiltonians are compared with each other
and with connection rules that are commonly used to match the wave
functions found from the bulk
\textbf{\textit{k}}$\cdot$\textbf{\textit{p}} Hamiltonians of two
adjacent materials. Electron and hole energy spectra in three
spherical QDHs: HgS/CdS, InAs/GaAs, and GaAs/AlAs are calculated
as a function of the quantum dot radius within the approximate
symmetrized and exact nonsymmetrized 8$\times$8 models. The
parameters of dissymmetry are shown to influence the energy levels
and the wave functions of an electron and a hole and,
consequently, the energies of both intraband and interband
transitions.
\end{abstract}


\section{Introduction}

The 4$\times$4 \textbf{\textit{k}}$\cdot$\textbf{\textit{p}} hole
Hamiltonian for the wave function envelopes (so called
effective-mass Hamiltonian), that takes into account mixing of the
light- and heavy-hole bands, was obtained in
Ref.~\onlinecite{Luttinger} using the perturbation theory. This
multiband Hamiltonian has been employed for description of the
hole states in bulk crystals\cite{Baldereschi} as well as in
low-dimensional structures, in particular, in free-standing
homogeneous quantum dots (QDs)\cite{Xia,Fomin}. The inclusion of
the mixing with the spin-orbit split-off hole band leads to the
6$\times$6 \textbf{\textit{k}}$\cdot$\textbf{\textit{p}}
Hamiltonian which has also been applied\cite{Grigoryan,Ekimov} to
QDs. To consider the nonparabolicity of the electron dispersion in
narrow- and medium-gap semiconductors, it is necessary to take
into account the coupling of the conduction and valence bands.
Using the \textbf{\textit{k}}$\cdot$\textbf{\textit{p}}
perturbation theory for bulk semiconductors with cubic lattice
symmetry, the 8$\times$8
\textbf{\textit{k}}$\cdot$\textbf{\textit{p}} model was developed
in Ref.~\onlinecite{Pidgeon}. This model explicitly includes eight
bands around the $\Gamma$ point of the Brillouin zone, namely,
electron, heavy-, light-, and spin-orbit split-off hole bands
(each of them is twice-degenerate due to the spin), and treats all
other bands as remote. Along with more simple models, the
8$\times$8 \textbf{\textit{k}}$\cdot$\textbf{\textit{p}}
Hamiltonian has been used to investigate different QDs (see, e.~g.
Refs.~\onlinecite{SercelV,Efros,Gashimzade,Sercel}).

Recently, one has begun to apply multiband effective-mass
Hamiltonians to investigate elastic, electronic, and optical
properties of multilayer nanostructures such as quantum-dot
heterostructures (QDHs): CdS/HgS\cite{Sercel},
InAs/GaAs\cite{Pryor,Stier},
GaAs/Al$_x$Ga$_{1-x}$As\cite{Shi,Hsieh}, and
CdS/HgS/CdS/H$_2$O\cite{Jaskolski,Pokatilov}. However, it should
be emphasized, that multiband
\textbf{\textit{k}}$\cdot$\textbf{\textit{p}} Hamiltonians are
derived for homogeneous bulk materials, i.e. under the assumption
that all effective-mass parameters are \emph{constant}. This is
important, because at a certain step of the derivation,
wavenumbers $\textbf{\textit{k}}$ are declared as
operators $\hat{\textbf{\textit{p}}}/\hbar$ that do not commute
with the functions of coordinates. But, at the heterointerfaces
of the multilayer nanostructures, there occurs an abrupt change of
effective-mass parameters from their values in one material to
those in the adjacent material. Inside a thin
transitional layer that contains the heterointerface, the
ordering of the differential operators and coordinate-dependent
effective-mass parameters in the multiband Hamiltonian
becomes crucial. In QDs with an infinitely high confining
potential for electrons and holes, all components of the wave
function vanish at the heterointerface, and there remains a
possibility of applying the bulk multiband
\textbf{\textit{k}}$\cdot$\textbf{\textit{p}} Hamiltonian
straightforwardly.
\cite{Xia,Fomin,Grigoryan,Ekimov,SercelV,Efros,Gashimzade,Sercel}
There are two ways to proceed from QDs to QDHs.

(i) The first way is to use an appropriate bulk multiband
Hamiltonian for each constituent material separately, and then to
match the obtained homogeneous solutions at the abrupt
heterojunctions applying the \emph{connection rules} (CRs) that
are usually obtained by imposing the continuity of the wave
function envelopes and of the normal to the heterointerface
component of the velocity.\cite{Sercel,Jaskolski} It should be
underlined that this way is heuristic and nonunique. In
Ref.~\onlinecite{Kisin} the general CRs, that even do not require
the continuity of the wave function envelopes, have been proposed
for planar heterostructures.

(ii) The second way (cf. Refs. \onlinecite{Burt1,Burt2,Burt3}) is
to derive a multiband Hamiltonian valid for the entire
heterostructure, including the heterointerfaces, and then, if
material parameters change abruptly at some interfaces, to find
the \emph{boundary conditions} (BCs) for the solutions of the
envelope function equation. To find these BCs, one should use the
multiband envelope function equation
$\left(\hat{H}-E\right)\Psi=0$ at any point of the
heterostructure, including the heterointerfaces, and integrate
this equation over the volume of an infinitely thin layer, which
includes the considered heterointerface. Thus, the BCs are derived
starting from the requirement of continuity of the components of
the wave function at the heterointerface.

One can always choose the CRs physically equivalent to the
BCs.\cite{Foreman0} The both approaches (i) and (ii) are usually used
when the wave function inside each layer of a heterostructure can
be found analytically, for example in planar or spherical
heterostructures. In case of an arbitrary shape of the
heterointerface, the approach (ii) can still be used because,
when the Hamiltonian is known for the entire heterostructure, one
can find an overall numerical solution of the Schr\"odinger
equation.

A commonly used heuristic method to obtain a multiband
effective-mass Hamiltonian for heterostructures uses
symmetrization\cite{Bastard,Lin,Baraff,Chao,Edwards} of the
corresponding \textbf{\textit{k}}$\cdot$\textbf{\textit{p}}
Hamiltonian. This method consists in the symmetrical arrangement
of the components of the momentum operator, that ensures the
hermicity of the resulting Hamiltonian. Namely
$\beta\,\hat{\textbf{\textit{p}}}\rightarrow\left(\beta({\bf
r})\,\hat{\textbf{\textit{p}}}+\hat{\textbf{\textit{p}}}\,\beta({\bf
r})\right)/2$ and
$\beta\,\hat{p}_i\,\hat{p}_j\rightarrow\left(\hat{p}_i\,\beta({\bf
r})\,\hat{p}_j+\hat{p}_j\,\beta({\bf r})\,\hat{p}_i\right)/2$,
where $\beta({\bf r})$ is a spatially varying effective-mass or
other material-dependent parameter which is usually considered a
piece-wise constant, because in each layer of a heterostructure it
has the value for a corresponding bulk material. The
symmetrization has been applied to QDHs in
Refs.~\onlinecite{Pryor,Stier,Pokatilov}. An essential fault of
the symmetrization is that it is not a {\it necessary} condition
for the multiband Hamiltonian to be hermitian. Besides that, as
will be seen below, some intrinsic properties of the
heterointerface, such as reducing the symmetry of the problem and
smoothing the abrupt change of the effective-mass parameters at a
heterojunction, are completely neglected in the symmetrized
Hamiltonian.

Burt has derived\cite{Burt1,Burt2,Burt3} the exact
envelope-function equations for a heterostructure. The order of
the components of the momentum operator arises as a part of that
derivation. This theory has been used by Foreman to explicitly
write the 6$\times$6 (Ref. \onlinecite{Foreman1}) and 8$\times$8
(Ref. \onlinecite{Foreman2}) effective-mass Hamiltonians for
planar heterostructures. General rules for constructing the
valence-band effective-mass Hamiltonians with a correct operator
ordering have been described in Ref.~\onlinecite{Dalen} for the
heterostructures with arbitrary crystallographic orientations. In
Ref.~\onlinecite{Mireles,Mireles2}, correct boundary conditions
for planar heterostructures with wurtzite symmetry have been
presented. Comparing the conduction- and valence-subband
dispersion of a planar quantum well, calculated using the BCs
following from the exact nonsymmetrized and from the symmetrized
effective-mass Hamiltonians, it has been shown that the former BCs
give physically reasonable results, while the latter BCs can
produce nonphysical solutions.\cite{Foreman1,Meney} More recently,
these two sets of BCs for a planar quantum well have been examined
within the tight-binding approach.\cite{Franceschi} The result of
the comparison allowed to give preference to the nonsymmetrized
model resulting from Burt's derivation of the envelope-function
Hamiltonian, which was shown to give reliable results even when
the well and barrier effective-mass parameters were very
dissimilar.

In the present paper the envelope-function representation of
Refs.~\onlinecite{Burt1,Burt2,Burt3} is used to construct the
nonsymmetrized 8-band Hamiltonian for an arbitrary 3-dimensional
heterostructure. As an application, the electronic structure of
two-layer HgS/CdS, InAs/GaAs, and GaAs/AlAs spherical QDHs is
investigated as a function of the dot radius.

It should be mentioned that the \emph{spurious solutions}
(``oscillating''\cite{Wang} states and ``gap''
states\cite{Sercel,Wang}) did not become apparent in the
aforementioned QDHs. However, such solutions may appear for a
different set of parameters.

The results of the calculation are compared with those obtained
from the symmetrized 8$\times$8 Hamiltonian. The rest of the paper
is organized as follows. In Sec.~II the derivation of the
nonsymmetrized 8-band Hamiltonian for a QDH is presented. The
corresponding radial Hamiltonian for a spherical QDH is obtained
in Sec.~III. In Sec.~IV the BCs for both symmetrized and
nonsymmetrized radial Hamiltonians are compared with each other
and with commonly used CRs. The results of the numerical
calculation for spherical QDHs are obtained and discussed in
Sec.~V. Conclusions are given in Sec.~VI. The 2$\times$2 electron
and 6$\times$6 hole energy-dependent nonsymmetrized Hamiltonians
for a QDH, as well as radial Hamiltonians and corresponding BCs
for a spherical QDH are found in Appendix~B from the
nonsymmetrized 8-band Hamiltonians.

\section{Nonsymmetrized 8-band Hamiltonian}

We begin our derivation with the nonsymmetrized 8-band
effective-mass Hamiltonian for a heterostructure, when the
spin-orbit coupling is ``turned-off''. In the Bloch function
basis $|S\rangle$, $|X\rangle$, $|Y\rangle$, $|Z\rangle$ this
Hamiltonian is represented in the following form\cite{Foreman2}
\begin{equation}\label{H44}
{\hskip-36pt} \hat{H}_4=\frac{\hbar^2}{2m_0}\left(
\begin{array}{cccc}
\varepsilon_c+\hat{\textbf{\textit{k}}}\alpha\hat{\textbf{\textit{k}}}&
\displaystyle\frac{i}{2}\left(v_1\hat{k}_x+\hat{k}_x v_2\right)&
\displaystyle\frac{i}{2}\left(v_1\hat{k}_y+\hat{k}_y v_2\right)&
\displaystyle\frac{i}{2}\left(v_1\hat{k}_z+\hat{k}_z
v_2\right)\\[8pt]
-\displaystyle\frac{i}{2}\left(v_2\hat{k}_x+\hat{k}_x v_1\right)&
\varepsilon_v^{\prime}-\hat{k}_x\beta_l\hat{k}_x
-\hat{\textbf{\textit{k}}}^{\perp}_x\beta_h\hat{\textbf{\textit{k}}}^{\perp}_x&
-3\left(\hat{k}_x\gamma_3^+\hat{k}_y+\hat{k}_y\gamma_3^-\hat{k}_x\right)&
-3\left(\hat{k}_x\gamma_3^+\hat{k}_z+\hat{k}_z\gamma_3^-\hat{k}_x\right)\\[8pt]
-\displaystyle\frac{i}{2}\left(v_2\hat{k}_y+\hat{k}_y v_1\right)&
-3\left(\hat{k}_x\gamma_3^-\hat{k}_y+\hat{k}_y\gamma_3^+\hat{k}_x\right)&
\varepsilon_v^{\prime}-\hat{k}_y\beta_l\hat{k}_y
-\hat{\textbf{\textit{k}}}^{\perp}_y\beta_h\hat{\textbf{\textit{k}}}^{\perp}_y&
-3\left(\hat{k}_y\gamma_3^+\hat{k}_z+\hat{k}_z\gamma_3^-\hat{k}_y\right)\\[8pt]
-\displaystyle\frac{i}{2}\left(v_2\hat{k}_z+\hat{k}_z v_1\right)&
-3\left(\hat{k}_x\gamma_3^-\hat{k}_z+\hat{k}_z\gamma_3^+\hat{k}_x\right)&
-3\left(\hat{k}_y\gamma_3^-\hat{k}_z+\hat{k}_z\gamma_3^+\hat{k}_y\right)&
\varepsilon_v^{\prime}-\hat{k}_z\beta_l\hat{k}_z
-\hat{\textbf{\textit{k}}}^{\perp}_z\beta_h\hat{\textbf{\textit{k}}}^{\perp}_z
\end{array}
\right),
\end{equation}
where $\varepsilon_v^{\prime}=\varepsilon_v-\delta/3$,\quad
$\hat{\textbf{\textit{k}}}=-i\nabla$,\quad
$\hat{\textbf{\textit{k}}}^{\perp}_{x,y,z}=\hat{\textbf{\textit{k}}}-
\hat{\textbf{\textit{k}}}_{x,y,z}$,\quad
$\beta_l=\gamma_1+4\gamma_2$,\quad $\beta_h=\gamma_1-2\gamma_2$,
\begin{equation}\label{xi}
v_1=v+\xi,\quad v_2=v-\xi,
\end{equation}
\begin{equation}\label{chi}
\gamma_3^+=\gamma_3+\chi,\quad \gamma_3^-=\gamma_3-\chi.
\end{equation}
$\xi$ and $\chi$ are called the dissymmetry parameters, because
when $\xi=0$ and $\chi=0$, the Hamiltonian~(\ref{H44}) becomes
symmetrical. The explicit form of the parameter
$\xi\equiv(v_1-v_2)/2$ follows from the formulae~(\ref{v1d}) and
(\ref{v2d}) in Appendix~A, and the parameter $\chi$ is determined
by Eq.~(\ref{chidef}).

When the spin-orbit coupling is ``turned-on'', the considered
8-band Hamiltonian is represented in the Bloch function basis
$|S\!\!\uparrow\rangle$, $|X\!\!\uparrow\rangle$,
$|Y\!\!\uparrow\rangle$, $|Z\!\!\uparrow\rangle$,
$|S\!\!\downarrow\rangle$, $|X\!\!\downarrow\rangle$,
$|Y\!\!\downarrow\rangle$, $|Z\!\!\downarrow\rangle$ as
\begin{equation}\label{H88}
\hat{H}_8=\left(
\begin{array}{cc}
\hat{H}_4&0\\
0&\hat{H}_4
\end{array}
\right)+H_{so},
\end{equation}
where $\hat{H}_4$ is defined by Eq.~(\ref{H44}) and the
spin-orbit Hamiltonian $H_{so}$ has the form\cite{Baraff}
\begin{equation}\label{Hso}
H_{so}=\frac{\Delta}{3}\left(
\begin{array}{cccccccc}
0&0&0&0&0&0&0&0\\[-5pt]
0&0&-i&0&0&0&0&1\\[-5pt]
0&i&0&0&0&0&0&-i\\[-5pt]
0&0&0&0&0&-1&i&0\\[-5pt]
0&0&0&0&0&0&0&0\\[-5pt]
0&0&0&-1&0&0&i&0\\[-5pt]
0&0&0&-i&0&-i&0&0\\[-5pt]
0&1&i&0&0&0&0&0
\end{array}
\right).
\end{equation}
In Eq.~(\ref{H44}), $m_0$ is the free-electron mass,
$E_c=\hbar^2\varepsilon_c/2m_0$ is the energy of the conduction
band (CB) minimum, $E_v=\hbar^2\varepsilon_v/2m_0$ is the energy
of the valence band (VB) maximum, $\Delta=\hbar^2\delta/2m_0$ is
the spin-orbit splitting of the VB, and $V=\hbar v/2m_0$ is the
Kane velocity ($V=-i\hbar\langle S|\hat{k}_z|Z\rangle/m_0$).
Contributions of remote bands to the hole effective masses are
written in terms of the ``modified'' Luttinger parameters
$\gamma_1=\gamma_1^L-E_p/3E_g$,\,\,
$\gamma_2=\gamma_2^L-E_p/6E_g$, and
$\gamma_3=\gamma_3^L-E_p/6E_g$, where $E_g=E_c-E_v$ is the energy
gap, $E_p=2m_0V^2$ is the Kane energy, and $\gamma_i^L (i=1,2,3)$ are the
Luttinger parameters of the VB. Parameter $\alpha$ can be
evaluated through the experimentally determined CB-mass $m_c$
using the relation
\begin{equation}\label{mc}
\frac{1}{m_c}=\frac{1}{m_0}\left(\alpha+\frac{E_p}{3}\left\lbrack\frac{2}{E_g}+
\frac{1}{E_g+\Delta}\right\rbrack\right).
\end{equation}
It is worth noting that all parameters entering the Hamiltonian (\ref{H88}) are
coordinate-dependent. In a heterostructure these parameters
abruptly change from their values in one material to the
corresponding values in the adjacent material, therefore they are
piecewise-constant functions of ${\bf r}$. Although not
symmetrical, the Hamiltonian $\hat{H}_8$ is hermitian as seen from
Eq.~(\ref{H88}). The parameters $\xi$ and $\chi$ (see Eqs.~(\ref{xi})
and (\ref{chi})) are responsible for the nonsymmetrical form of
the Hamiltonian (\ref{H88}). The symmetrized 8-band Hamiltonian
can be obtained, therefore, taking $\xi=0$ and $\chi=0$.

In Eq.~(\ref{chi}), $\gamma_3^+$ is the contribution to $\gamma_3$
from the $\Gamma_1$ and $\Gamma_{12}$ remote bands, while
$\gamma_3^-$ is the contribution to $\gamma_3$ from the $\Gamma_{15}$
and $\Gamma_{25}$ remote bands.\cite{Foreman2} Neglecting small
contributions from the $\Gamma_{25}$ remote bands, the parameter
$\chi({\bf r})$ is determined as\cite{Foreman2}
\begin{equation}\label{chidef}
\chi({\bf r})=(2\gamma_2({\bf r})+3\gamma_3({\bf
r})-\gamma_1({\bf r})-1)/3,
\end{equation}
i.e. it is explicitly defined by the effective-mass parameters of
the bulk model. It is seen from Eq.~(\ref{chidef}) that in a
homogeneous medium when $\gamma_i({\bf r})$ are constants,
$\chi({\bf r})$ is also a constant, and therefore, it cancels from
Eq.~(\ref{H44}). Consequently, $\chi({\bf r})$ is a specific
function of a heterostructure, which gives a nonzero contribution
to the Hamiltonian only at the heterointerfaces. The value of this
contribution at the point ${\bf r}_0$ of a heterointerface is
proportional to $\chi({\bf r}_0+{\bf e}_n)-\chi({\bf r}_0-{\bf
e}_n)$, where ${\bf e}_n$ is an infinitesimally small vector,
normal to the heterointerface at the point ${\bf r}_0$. Parameters
$v_1({\bf r})$ and $v_2({\bf r})$ of the Hamiltonian~(\ref{H44}),
which can be obtained from the general effective mass equations as
derived by Burt\cite{Burt1}, are given in Appendix~A. In the
definition~(\ref{xi}), the functions $v_1({\bf r})$ and $v_2({\bf
r})$ are subdivided into the symmetric $v({\bf r})$ and
antisymmetric $\xi({\bf r})$ parts, where $\xi({\bf r})$, like
$\chi({\bf r})$ above, is a specific parameter of a
heterostructure. In general, $\xi({\bf r})$ is a
piecewise-constant function of ${\bf r}$. The necessary and
sufficient condition for $\xi({\bf r})$ to give a nonzero
contribution to the 8$\times$8 Hamiltonian only at the
heterointerfaces, and to become a constant in the homogeneous
medium, simultaneously with $v({\bf r})$, is
\begin{equation}\label{cxi}
\xi({\bf r})=c_\xi\,v({\bf r}),
\end{equation}
where the coefficient of proportionality $c_\xi$ is constant
over the entire heterostructure. Eq.~(\ref{cxi}) is the
general form of $\xi({\bf r})$ only for a two-layer
heterostructure. For an $N$-layer heterostructure, there can be
$N-1$ independent constants -- one for each heterointerface. Each
constant for a given heterointerface can be found
experimentally considering a two-layer heterostructure (see Sec.~V).

In order to diagonalize the
spin-orbit Hamiltonian $H_{so}$, it is convenient to carry out a
unitary transformation of the Bloch function basis
$|S\!\!\uparrow\rangle$, $|X\!\!\uparrow\rangle$,
$|Y\!\!\uparrow\rangle$, $|Z\!\!\uparrow\rangle$,
$|S\!\!\downarrow\rangle$, $|X\!\!\downarrow\rangle$,
$|Y\!\!\downarrow\rangle$, $|Z\!\!\downarrow\rangle$ into the
following Bloch function basis\cite{Efros}
\begin{equation}\label{bas.c}
\begin{array}{rcl}
u_{1/2,1/2}^c&=&|S\!\!\uparrow\rangle,\\[10pt]
u_{1/2,-1/2}^c&=&|S\!\!\downarrow\rangle,
\end{array}
\end{equation}
\begin{equation}\label{bas.v}
\begin{array}{rcl}
u_{3/2,3/2}^v&=&\displaystyle{\frac{1}{\sqrt{2}}
\Bigl(|X\!\!\uparrow\rangle+i|Y\!\!\uparrow\rangle\Bigr)},\\[10pt]
u_{3/2,1/2}^v&=&\displaystyle{\frac{i}{\sqrt{6}}\Bigl(|X\!\!\downarrow\rangle+
i|Y\!\!\downarrow\rangle-2|Z\!\!\uparrow\rangle\Bigr)},\\[10pt]
u_{3/2,-1/2}^v&=&\displaystyle{\frac{1}{\sqrt{6}}\Bigl(|X\!\!\uparrow\rangle-
i|Y\!\!\uparrow\rangle+2|Z\!\!\downarrow\rangle\Bigr)},\\[10pt]
u_{3/2,-3/2}^v&=&\displaystyle{\frac{i}{\sqrt{2}}
\Bigl(|X\!\!\downarrow\rangle-i|Y\!\!\downarrow\rangle\Bigr)},\\[10pt]
u_{1/2,1/2}^v&=&\displaystyle{\frac{1}{\sqrt{3}}\Bigl(|X\!\!\downarrow\rangle+
i|Y\!\!\downarrow\rangle+|Z\!\!\uparrow\rangle\Bigr)},\\[10pt]
u_{1/2,-1/2}^v&=&\displaystyle{\frac{-i}{\sqrt{3}}\Bigl(|X\!\!\uparrow\rangle-
i|Y\!\!\uparrow\rangle-|Z\!\!\downarrow\rangle\Bigr)},
\end{array}
\end{equation}
where $u_{J,\mu}^c$ and $u_{J,\mu}^v$ are the Bloch functions of
the conduction and valence bands, $J$ is the Bloch function
angular momentum, and $\mu\equiv J_z$ is its $z$-component. The
8-band Hamiltonian $\hat{H}$ in the new basis can be obtained now
after a unitary transformation
\begin{equation}\label{transf}
\hat{H}=U^*\,\hat{H}_8\,U^T,
\end{equation}
where $U$ is the matrix of the transformation from the basis
$\{|S\!\!\uparrow\rangle$, $|X\!\!\uparrow\rangle$,
$|Y\!\!\uparrow\rangle$, $|Z\!\!\uparrow\rangle$,
$|S\!\!\downarrow\rangle$, $|X\!\!\downarrow\rangle$,
$|Y\!\!\downarrow\rangle$, $|Z\!\!\downarrow\rangle\}$ to the new
basis $\{u_{1/2,1/2}^c$, $u_{1/2,-1/2}^c$, $u_{3/2,3/2}^v$,
$u_{3/2,1/2}^v$, $u_{3/2,-1/2}^v$, $u_{3/2,-3/2}^v$,
$u_{1/2,1/2}^v$, $u_{1/2,-1/2}^v\}$. $U$ is defined by
Eqs.~(\ref{bas.c}) and (\ref{bas.v}). Performing the
transformation~(\ref{transf}), one obtains:
\begin{equation}\label{H}
{\hskip-32pt} \hat{H}=\frac{\hbar^2}{2m_0}\left(
\begin{array}{cccccccc}
\varepsilon_c+T & 0 & i V_1 & \displaystyle{\sqrt{\frac{2}{3}}V_0}
& \displaystyle{\frac{i}{\sqrt{3}}V_{-1}} & 0 &
\displaystyle{\frac{i}{\sqrt{3}}V_0} &
\displaystyle{\sqrt{\frac{2}{3}}V_{-1}}\\[10pt]
0 & \varepsilon_c+T & 0 & \displaystyle{\frac{-1}{\sqrt{3}}V_1} &
\displaystyle{i\sqrt{\frac{2}{3}}V_0} & -V_{-1} &
\displaystyle{i\sqrt{\frac{2}{3}}V_1} &
\displaystyle{\frac{-1}{\sqrt{3}}V_0}\\[10pt]
-iV^\dag_1 & 0 & \varepsilon_v-P-Q & -S & -R & 0 &
\displaystyle{\frac{-i}{\sqrt{2}}S} & i\sqrt{2}R\\[10pt]
\displaystyle{\sqrt{\frac{2}{3}}V^\dag_0} &
\displaystyle{\frac{-1}{\sqrt{3}}V^\dag_1} & -S^\dag &
\varepsilon_v-P+Q & -C & -R & i\sqrt{2}Q &
\displaystyle{-i\sqrt{\frac{3}{2}}\Sigma}\\[10pt]
\displaystyle{\frac{-i}{\sqrt{3}}V^\dag_{-1}} &
\displaystyle{-i\sqrt{\frac{2}{3}}V^\dag_0} & -R^\dag & -C^\dag &
\varepsilon_v-P^*+Q^* & S^T &
\displaystyle{i\sqrt{\frac{3}{2}}\Sigma^*} &
i\sqrt{2}Q^*\\[10pt]
0 & -V^\dag_{-1} & 0 & -R^\dag & S^* & \varepsilon_v-P^*-Q^*
& i\sqrt{2}R^\dag & \displaystyle{\frac{i}{\sqrt{2}}S^*}\\[10pt]
\displaystyle{\frac{-i}{\sqrt{3}}V^\dag_0} &
\displaystyle{-i\sqrt{\frac{2}{3}}V^\dag_1} &
\displaystyle{\frac{i}{\sqrt{2}}S^\dag} & -i\sqrt{2}Q &
\displaystyle{-i\sqrt{\frac{3}{2}}\Sigma^T} &
-i\sqrt{2}R & \varepsilon^{\prime \prime}_v-P & C\\[10pt]
\displaystyle{\sqrt{\frac{2}{3}}V^\dag_{-1}} &
\displaystyle{\frac{-1}{\sqrt{3}}V^\dag_0} & -i\sqrt{2}R^\dag &
\displaystyle{i\sqrt{\frac{3}{2}}\Sigma^\dag} & -i\sqrt{2}Q^* &
\displaystyle{\frac{-i}{\sqrt{2}}S^T} & C^\dag &
\varepsilon^{\prime \prime}_v-P^*
\end{array}
\right),
\end{equation}
where $\varepsilon^{\prime \prime}_v=\varepsilon_v-\delta$,
$$
\hat{k}_+=\frac{\hat{k}_x+i\,\hat{k}_y}{\sqrt{2}},\quad
\hat{k}_-=\frac{\hat{k}_x-i\,\hat{k}_y}{\sqrt{2}},
$$
$$
V_1=\frac{1}{2}\left(v_1\hat{k}_++\hat{k}_+ v_2\right),\quad
V_{-1}=\frac{1}{2}\left(v_1\hat{k}_-+\hat{k}_- v_2\right),
$$
$$
V_0=\frac{1}{2}\left(v_1\hat{k}_z+\hat{k}_z v_2\right),\quad
T=\hat{k}_+\alpha\hat{k}_-+\hat{k}_-\alpha\hat{k}_++\hat{k}_z\alpha\hat{k}_z,
$$
$$
P=\hat{k}_+(\gamma_1-2\chi)\hat{k}_-+\hat{k}_-(\gamma_1+2\chi)\hat{k}_+
+\hat{k}_z\gamma_1\hat{k}_z,
$$
$$
Q=\hat{k}_+(\gamma_2-\chi)\hat{k}_-+\hat{k}_-(\gamma_2+\chi)\hat{k}_+
-2\hat{k}_z\gamma_2\hat{k}_z,
$$
$$
R=\sqrt{3}\left(\hat{k}_+(\gamma_2-\gamma_3)\hat{k}_+
+\hat{k}_-(\gamma_2+\gamma_3)\hat{k}_-\right),
$$
$$
S=-i\sqrt{6}\left(\hat{k}_-(\gamma_3+\chi)\hat{k}_z
+\hat{k}_z(\gamma_3-\chi)\hat{k}_-\right),
$$
$$
\Sigma=-i\sqrt{6}\left(\hat{k}_-(\gamma_3-\frac{\chi}{3})\hat{k}_z
+\hat{k}_z(\gamma_3+\frac{\chi}{3})\hat{k}_-\right),
$$
\begin{equation}\label{C}
C=-i\,2\sqrt{2}\left(\hat{k}_-\chi\hat{k}_z-\hat{k}_z\chi\hat{k}_-\right).
\end{equation}
In Eq.~(\ref{H}), daggers (\dag) denote the hermitian conjugation,
i.e. $A^\dag\equiv\left(A^T\right)^*$ (it is important to note,
that $\left(v_{1,2}\,\hat{k}_\pm\right)^\dag=\hat{k}_\mp\,v_{1,2}$
and $\left(v_{1,2}\,\hat{k}_z\right)^\dag=\hat{k}_z\,v_{1,2}$).
Unlike the bulk 8$\times$8
\textbf{\textit{k}}$\cdot$\textbf{\textit{p}}
Hamiltonian\cite{Efros} where the matrix element $C$ is zero, in
the Hamiltonian (\ref{H}) the Bloch functions $u_{3/2,1/2}^v$ and
$u_{3/2,-1/2}^v$ are coupled with each other, as well as the
functions $u_{1/2,1/2}^v$ and $u_{1/2,-1/2}^v$. As seen from
Eq.~(\ref{C}), this coupling arises because of the dissymmetry
parameter $\chi$, which can reduce, in this way, the symmetry of
the problem. The Hamiltonian so obtained can be used to
investigate electronic properties of quantum-well, quantum-wire,
and quantum-dot heterostructures.

\section{8-band Hamiltonian for a spherical QDH}

To study the electronic structure of spherical QDHs, the spherical
approximation\cite{Baldereschi} (i.e.
$\gamma_2^L=\gamma_3^L\equiv\gamma^L$) can be applied. If we take
$\gamma^L=(2\gamma_2^L+3\gamma_3^L)/5$, then the quantum states so
obtained are correct to the first order of the perturbation
theory. Using the relations between the Luttinger parameters
$\gamma_i^L$ and the ``modified'' Luttinger parameters $\gamma_i$
we have
\begin{equation}\label{gamma8}
\gamma=(2\gamma_2+3\gamma_3)/5,
\end{equation}
and according to Eq.~(\ref{chidef})
\begin{equation}\label{chi8}
\chi=(5\gamma-\gamma_1-1)/3.
\end{equation}
In spherical QDHs, where all effective-mass parameters depend
only on the absolute value $r$ of the radius-vector, electron and
hole states are eigenfunctions of the total angular momentum $j$
and its $z$-component $m\equiv j_z$. Therefore, the electron or
hole wave function can be written as a linear expansion in the
eight Bloch functions $u_{J,\mu}^{c(v)}$:
\begin{equation}\label{wf}
\Psi_{j,m}({\bf r})=\sum\limits_{J,\mu} F^{c;j,m}_{J,\mu}({\bf
r})\,u_{J,\mu}^c + \sum\limits_{J,\mu} F^{v;j,m}_{J,\mu}({\bf
r})\,u_{J,\mu}^v,
\end{equation}
where the envelope functions $F^{c(v);j,m}_{J,\mu}({\bf r})$ are
defined in the chosen Bloch function basis (\ref{bas.c}),
(\ref{bas.v}) as
\begin{eqnarray}\label{envelopes}
F^{c;j,m}_{1/2,\mu}({\bf r})&=&\sum\limits_{l,\lambda}
C_{1/2,\mu;l,\lambda}^{j,m}\,R^{c;j}_{1/2,l}(r)\,
Y_{l,\lambda}(\theta,\phi),\nonumber\\
F^{v;j,m}_{3/2,\mu}({\bf r})&=&i^{\mu-3/2}\sum\limits_{l,\lambda}
C_{3/2,\mu;l,\lambda}^{j,m}\,R^{v;j}_{3/2,l}(r)\,
Y_{l,\lambda}(\theta,\phi),\\
F^{v;j,m}_{1/2,\mu}({\bf r})&=&i^{1/2-\mu}\sum\limits_{l,\lambda}
C_{1/2,\mu;l,\lambda}^{j,m}\,R^{v;j}_{1/2,l}(r)\,
Y_{l,\lambda}(\theta,\phi).\nonumber
\end{eqnarray}
Here, $R^{c(v);j}_{J,l}(r)$ are the radial envelope functions,
$C_{J,\mu;l,\lambda}^{j,m}$ are the Clebsch-Gordan coefficients,
and $Y_{l,\lambda}(\theta,\phi)$ are the spherical harmonics.
Noting that in the matrix representation of the Hamiltonian
(\ref{H}) $u^c_{1/2,1/2}=\Big(1\quad 0\quad 0\quad 0\quad 0\quad
0\quad 0\quad 0\Big)^T$, \dots, $u^v_{1/2,-1/2}=\Big(0\quad 0\quad
0\quad 0\quad 0\quad 0\quad 0\quad 1\Big)^T$ one can rearrange
Eq.~(\ref{wf}) into the form
\begin{eqnarray}\label{wf2}
\Psi_{j,m}({\bf r})&=&
\sum\limits_{l=j-1/2}^{j+1/2}R^{c;j}_{1/2,l}(r)\,\,
{\mathcal Y}^{c;j,m}_{1/2,l}(\theta,\phi)\nonumber\\
&+&\sum\limits_{l=j-3/2}^{j+3/2}R^{v;j}_{3/2,l}(r)\,\,
{\mathcal Y}^{v;j,m}_{3/2,l}(\theta,\phi)\\
&+&\sum\limits_{l=j-1/2}^{j+1/2}R^{v;j}_{1/2,l}(r)\,\, {\mathcal
Y}^{v;j,m}_{1/2,l}(\theta,\phi)\nonumber,
\end{eqnarray}
where the 8$\times$8 matrices ${\mathcal
Y}^{c(v);j,m}_{J,l}(\theta,\phi)$ next to the eight radial
envelope functions $R^{c(v);j}_{J,l}(r)$, for a given $j$, can be
found by comparing Eq.~(\ref{wf2}) with Eqs.~(\ref{wf}) and
(\ref{envelopes}). Now, integrating over the angular variables
$\theta$ and $\phi$, it is possible to obtain the radial
Hamiltonian
\begin{equation}\label{Hint}
\hat{{\mathcal H}}_j=\int \left({\mathcal
Y}^{b^\prime;j,m}_{J^\prime,l^\prime}(\theta,\phi)\right)^\dag
\hat{H}\,\, {\mathcal Y}^{b;j,m}_{J,l}(\theta,\phi)\,\,d\Omega,
\end{equation}
corresponding to the radial Schr\"odinger equation
\begin{equation}\label{Schr}
\sum\limits_{b,J,l}\hat{{\mathcal H}}_j\,\,R^{b;j}_{J,l}(r)
=E_j\,\,R^{b;j}_{J,l}(r),
\end{equation}
where $E_j$ is the electron or hole eigenenergy to be determined,
$b=c\mbox{ or }v$. The Hamiltonian (\ref{Hint}) does not depend on
$m$, because within the spherical approximation the energy
spectrum is degenerate with respect to the $z$-component of the
total momentum.

After some algebra, we derive
the following relations for the spherical harmonics:
\begin{eqnarray}
\hat{k}_+Y_{l,\lambda}(\theta,\phi)&=
&C_{l+1,\lambda+1;1,-1}^{l,\lambda}\,B_l^+
\,Y_{l+1,\lambda+1}(\theta,\phi)\nonumber\\
&+&C_{l-1,\lambda+1;1,-1}^{l,\lambda}\,B_l^-
\,Y_{l-1,\lambda+1}(\theta,\phi)\nonumber,
\end{eqnarray}
\begin{eqnarray}
\hat{k}_-Y_{l,\lambda}(\theta,\phi)&=
&-C_{l+1,\lambda-1;1,1}^{l,\lambda}\,B_l^+
\,Y_{l+1,\lambda-1}(\theta,\phi)\nonumber\\
&&-C_{l-1,\lambda-1;1,1}^{l,\lambda}\,B_l^-
\,Y_{l-1,\lambda-1}(\theta,\phi)\nonumber,
\end{eqnarray}
\begin{eqnarray}\label{kY}
\hat{k}_zY_{l,\lambda}(\theta,\phi)&=
&C_{l+1,\lambda;1,0}^{l,\lambda}\,B_l^+
\,Y_{l+1,\lambda}(\theta,\phi)\nonumber\\
&+&C_{l-1,\lambda;1,0}^{l,\lambda}\,B_l^-
\,Y_{l-1,\lambda}(\theta,\phi),
\end{eqnarray}
where
$$
B_l^+=-i\sqrt{\frac{l+1}{2l+1}}\,A_l^{(1)},\quad
B_l^-=-i\sqrt{\frac{l}{2l+1}}\,A_l^{(-1)},
$$
\begin{equation}\label{A}
A_l^{(p)}=-p\,\frac{\partial}{\partial r}+\frac{l+1/2-p/2}{r}.
\end{equation}
Using the relations (\ref{kY}), the radial Hamiltonian
(\ref{Hint}) can be obtained in an explicit form. If we choose the
following order of the radial functions: $R^{c;j}_{1/2,j-1/2}$,
$R^{v;j}_{3/2,j+1/2}$, $R^{v;j}_{3/2,j-3/2}$,
$R^{v;j}_{1/2,j+1/2}$, $R^{c;j}_{1/2,j+1/2}$,
$R^{v;j}_{3/2,j-1/2}$, $R^{v;j}_{3/2,j+3/2}$,
$R^{v;j}_{1/2,j-1/2}$, then the 8$\times$8 matrix of the
Hamiltonian $\hat{{\mathcal H}}_j$ takes the form
\begin{equation}\label{Hr8}
\hat{{\mathcal H}}_j=\left(
\begin{array}{cc}
\hat{{\mathcal H}}_j^{(1)}&0\\
0&\hat{{\mathcal H}}_j^{(-1)}
\end{array}
\right).
\end{equation}
Here, $\hat{{\mathcal H}}_j^{(1)}$ is the 4$\times$4 Hamiltonian
of the ``even'' states and $\hat{{\mathcal H}}_j^{(-1)}$ is the
4$\times$4 Hamiltonian of the ``odd'' states. It is seen, that the
parity $p$ ($p=1$ for even states and $p=-1$ for odd states) is
conserved in the spherical approximation, even when the
Hamiltonian is {\it not} symmetrized. The obtained radial Hamiltonian
$\hat{{\mathcal H}}_j^{(p)}$ for the radial functions
$R^{c;j}_{1/2,j-p/2}$, $R^{v;j}_{3/2,j+p/2}$,
$R^{v;j}_{3/2,j-3p/2}$, and $R^{v;j}_{1/2,j+p/2}$ has the form
\begin{equation}\label{Hjp}
{\hskip-9pt} \hat{{\mathcal H}}_j^{(p)}=\frac{\hbar^2}{2m_0}\left(
\begin{array}{cccc}
\varepsilon_c-{\mathcal T}_{j-p/2}& a_j^p\,{\mathcal
A}_{j+p/2}^{(-p)}& b_j^p\,{\mathcal A}_{j-3p/2}^{(p)}&
p\,\sqrt{2}\,{\mathcal A}_{j+p/2}^{(-p)}\\[8pt]
a_j^p\,{\mathcal B}_{j-p/2}^{(p)}& \varepsilon_v+{\mathcal
P}_{j+p/2}^{(-p)}-c_j^p\,{\mathcal Q}_{j+p/2}^{(-p)}&
a_j^p\,b_j^p\,{\mathcal R}_{j-3p/2}^{(p)}&
p\,\sqrt{2}\,a_j^p\,{\mathcal Q}_{j+p/2}^{(-p)}\\[8pt]
b_j^p\,{\mathcal B}_{j-p/2}^{(-p)}& a_j^p\,b_j^p\,{\mathcal
R}_{j+p/2}^{(-p)}& \varepsilon_v+{\mathcal
P}_{j-3p/2}^{(p)}+c_j^p\,{\mathcal Q}_{j-3p/2}^{(p)}&
p\,\sqrt{2}\,b_j^p\,{\mathcal R}_{j+p/2}^{(-p)}\\[8pt]
p\,\sqrt{2}\,{\mathcal B}_{j-p/2}^{(p)}&
p\,\sqrt{2}\,a_j^p\,{\mathcal Q}_{j+p/2}^{(-p)}&
p\,\sqrt{2}\,b_j^p\,{\mathcal R}_{j-3p/2}^{(p)}&
\varepsilon_v-\delta+{\mathcal P}_{j+p/2}^{(-p)}
\end{array}
\right),
\end{equation}
where $a_j^p=\sqrt{1+3\eta_j^p}$,\quad
$b_j^p=\sqrt{3(1-\eta_j^p)}$,\quad $c_j^p=1-3\eta_j^p$,\quad
$\eta_j^p=p/(2j+1-p)$,
\begin{eqnarray}\label{Rlp}
{\mathcal A}_{l}^{(p)}&=&\frac{1}{2\sqrt{6}}\left( v_1
A_{l}^{(p)}+A_{l}^{(p)} v_2\right),\nonumber\\[10pt]
{\mathcal B}_{l}^{(p)}&=&\frac{1}{2\sqrt{6}}\left( v_2
A_{l}^{(p)}+A_{l}^{(p)} v_1\right),\\[10pt]
{\mathcal
R}_{l}^{(p)}&=&-A_{l+p}^{(p)}\,\gamma\,A_{l}^{(p)}.\nonumber
\end{eqnarray}
Introducing the operator
\begin{equation}\label{Delta}
\Delta_l^{(p)}(\beta)=-A_{l+p}^{(-p)}\,\beta\,A_{l}^{(p)},
\end{equation}
we can represent ${\mathcal T}_{l}$, ${\mathcal P}_{l}^{(p)}$,
and ${\mathcal Q}_{l}^{(p)}$ as
\begin{eqnarray}\label{TPQ}
{\mathcal T}_{l}&=&\frac{(l+1)\Delta_l^{(1)}(\alpha)
+l\,\Delta_l^{(-1)}(\alpha)}{2l+1},\nonumber\\[10pt]
{\mathcal P}_{l}^{(p)}&=&\frac{(l+1)\Delta_l^{(1)}(\gamma_1-2\chi)
+l\,\Delta_l^{(-1)}(\gamma_1-2\chi)}{2l+1}+\Delta_l^{(p)}(2\chi),\\[10pt]
{\mathcal Q}_{l}^{(p)}&=&\frac{(l-1/2)\Delta_l^{(1)}(\gamma-\chi)
+(l+3/2)\Delta_l^{(-1)}(\gamma-\chi)}{2l+1}+\Delta_l^{(p)}(\chi).\nonumber
\end{eqnarray}
Inside each spherical layer, the radial Hamiltonian (\ref{Hjp})
for a spherical QDH coincides with the bulk radial Hamiltonian
from Ref.~\onlinecite{Efros} for a spherical QD, when the
following denotations for the radial functions are used:
\begin{equation}\label{otherR}
\begin{array}{rclrcl}
R^{c;j}_{1/2,j-1/2}&=&R^+_{c,j},&R^{c;j}_{1/2,j+1/2}&=&-R^-_{c,j},\\[8pt]
R^{v;j}_{3/2,j+1/2}&=&R^+_{h1,j},&R^{v;j}_{3/2,j-1/2}&=&R^-_{h1,j},\\[8pt]
R^{v;j}_{3/2,j-3/2}&=&-R^+_{h2,j},&R^{v;j}_{3/2,j+3/2}&=&-R^-_{h2,j},\\[8pt]
R^{v;j}_{1/2,j+1/2}&=&R^+_{s,j},&R^{v;j}_{1/2,j-1/2}&=&R^-_{s,j}.
\end{array}
\end{equation}
Therefore, in order to find the radial wave functions $R^{c(v);j}_{J,l}$
inside each spherical layer, one can use the same technique as in
Ref.~\onlinecite{Efros}. When the wave functions inside each
spherical layer are known, the BCs should be applied to match the
wave functions from two adjacent layers.

\section{Boundary conditions for a spherical QDH}

When considering the multiband models for planar
heterostructures, the BCs for the wave function are often obtained
by integrating the Schr\"odinger equation across the
heterointerface and assuming the continuity of the wave-function
envelopes.\cite{BenDaniel,Bastard,Burt1}. The resulting BCs are
of the following form:
\begin{equation}\label{BCpl}
\left.\Psi_A \right|_{z=-0}=\left.\Psi_B \right|_{z=+0},\quad
\left.\hat{{\mathcal J}}_z\,\Psi_A \right|_{z=-0}=
\left.\hat{{\mathcal J}}_z\,\Psi_B \right|_{z=+0},
\end{equation}
where $A$ and $B$ are two materials separated by the
heterointerface $z=0$, and $\hat{{\mathcal J}}_z$ is the normal to
the heterointerface component of the current operator. The
aforementioned integration is actually justified only for Burt's
envelope-function equations, because only these have been shown to
be valid at the heterointerface. Analogously to the case of planar
heterostructures, for spherical QDHs one integrates the radial
Schr\"odinger equation
\begin{equation}\label{SchrE}
\left(\hat{{\mathcal
H}}_j^{(p)}-E_j^{(p)}\right)\,R_j^{(p)}=0,\,\,
R_j^{(p)}=\left(\begin{array}{c}
R^{c;j}_{1/2,j-p/2}\\[8pt]
R^{v;j}_{3/2,j+p/2}\\[8pt]
R^{v;j}_{3/2,j-3p/2}\\[8pt]
R^{v;j}_{1/2,j+p/2}
\end{array}\right),
\end{equation}
where $\hat{{\mathcal H}}_j^{(p)}$ is defined by Eq.~(\ref{Hjp})
and $E_j^{(p)}$ is the eigenenergy. This integration is carried
out across the point $r=a$, where $r=a$ is the spherical
heterointerface that separates two materials: $A$ (at $r < a$) and
$B$ (at $r > a
$).
Including the continuity of the radial wave function $R_j^{(p)}$,
the required BCs have the form:
\begin{equation}\label{new}
\begin{array}{rcl}
\left.\left(R_{j}^{(p)}\right)_A\right|_{r=a-0}&=&
\left.\left(R_{j}^{(p)}\right)_B\right|_{r=a+0},\\[10pt]
\left.\hat{{\mathcal
J}}_j^{(p)}\,\left(R_{j}^{(p)}\right)_A\right|_{r=a-0}&=&
\left.\hat{{\mathcal
J}}_j^{(p)}\,\left(R_{j}^{(p)}\right)_B\right|_{r=a+0}.
\end{array}
\end{equation}
Here the radial component of the current operator $\hat{{\mathcal
J}}_j^{(p)}$ is obtained from the radial Hamiltonian
$\hat{{\mathcal H}}_j^{(p)}$ using the following procedure. In
those terms of the Hamiltonian (\ref{Hjp}) that contain the
operator $A_l^{(p)}$, the utmost left-hand-side $A_l^{(p)}$ is
replaced by $-p$ (in conformity with Eq.~(\ref{A})); the rest of
the terms are set to zero; the result is multiplied by $2i/\hbar$.
Thus we find
\begin{equation}\label{BC}
{\hskip-38pt} \hat{{\mathcal J}}_j^{(p)}=\frac{i\hbar}{m_0}\left(
\begin{array}{cccc}
-\alpha\,\displaystyle\frac{\partial}{\partial r}&
\displaystyle\frac{p}{2\sqrt{6}}\,a_j^p\,(v-\xi)&
\displaystyle\frac{-p}{2\sqrt{6}}\,b_j^p\,(v-\xi)&
\displaystyle\frac{1}{2\sqrt{3}}\,(v-\xi)\\[10pt]
\displaystyle\frac{-p}{2\sqrt{6}}\,a_j^p\,(v+\xi)&
\gamma_1\displaystyle\frac{\partial}{\partial r}
-c_j^p\,\gamma\,D_r+f_{-8}^{j,p}\,\displaystyle\frac{\chi}{r}&
p\,a_j^p\,b_j^p\,\gamma\,A_{j-3p/2}^{(p)}&
p\,\sqrt{2}\,a_j^p\left(\gamma\,D_r
+f_{1}^{j,p}\,\displaystyle\frac{\chi}{r}\right)\\[10pt]
\displaystyle\frac{p}{2\sqrt{6}}\,b_j^p\,(v+\xi)&
-p\,a_j^p\,b_j^p\,\gamma\,A_{j+p/2}^{(-p)}&
\gamma_1\displaystyle\frac{\partial}{\partial r}
+c_j^p\,\gamma\,D_r-3f_{4}^{j,p}\,\displaystyle\frac{\chi}{r}&
-\sqrt{2}\,b_j^p\,\gamma\,A_{j+p/2}^{(-p)}\\[10pt]
\displaystyle\frac{-1}{2\sqrt{3}}\,(v+\xi)&
p\,\sqrt{2}\,a_j^p\left(\gamma\,D_r
+f_{1}^{j,p}\,\displaystyle\frac{\chi}{r}\right)&
\sqrt{2}\,b_j^p\,\gamma\,A_{j-3p/2}^{(p)}&
\gamma_1\displaystyle\frac{\partial}{\partial
r}+2f_{-2}^{j,p}\,\displaystyle\frac{\chi}{r}
\end{array}
\right),
\end{equation}
where
\begin{equation}\label{Dr}
D_r=\frac{\partial}{\partial r}+\frac{3/2}{r},\quad
f_{n}^{j,p}=p\,(j+1/2-n\,p/2).
\end{equation}
While the radial Hamiltonian $\hat{{\mathcal H}}_j^{(p)}$ is
hermitian, the radial component of the current operator
$\hat{{\mathcal J}}_j^{(p)}$ is not (as seen from Eq.~(\ref{BC})).

It is important to compare the obtained BCs (\ref{new}),
(\ref{BC}) with the commonly used CRs that the wave function and
the normal component of the velocity are continuous at the
heterointerface\cite{Sercel,Jaskolski}. The velocity operator
$\hat{\textbf{V}}$ is defined as
\begin{equation}\label{V}
\hat{\textbf{V}}=\frac{i}{\hbar}\left[\hat{H},\,\textbf{r}\right]\equiv
\frac{1}{\hbar}\,\frac{\partial \hat{H}}{\partial
\hat{\textbf{\textit{k}}}},
\end{equation}
where the Hamiltonian $\hat{H}$ has been determined earlier by
Eq.~(\ref{H}). Therefore, the normal component of the velocity
operator is obtained as follows:
\begin{equation}\label{vr}
\hat{V}_r\equiv\frac{1}{\hbar}\,\frac{\textbf{r}}{r}\,\frac{\partial
\hat{H}}{\partial \hat{\textbf{\textit{k}}}}=\frac{1}{\hbar}\left(
n_+\,\frac{\partial \hat{H}}{\partial \hat{k}_+}+
n_-\,\frac{\partial \hat{H}}{\partial \hat{k}_-}+
n_z\,\frac{\partial \hat{H}}{\partial \hat{k}_z}\right),
\end{equation}
where
\begin{equation}\label{n}
n_+=\frac{x+i\,y}{r \sqrt{2}},\quad n_-=\frac{x-i\,y}{r
\sqrt{2}},\quad n_z=\frac{z}{r}.
\end{equation}
The differentiation of the Hamiltonian can be realized in such
a way that
\begin{equation}\label{Vel}
\hat{V}_r=\hat{V}_r^L+\hat{V}_r^R,
\end{equation}
where
\begin{equation}\label{vrLR}
\hat{V}_r^{L(R)}=\frac{1}{\hbar}\left( n_+\,\frac{\partial
\hat{H}^{L(R)}}{\partial \hat{k}_+}+ n_-\,\frac{\partial
\hat{H}^{L(R)}}{\partial \hat{k}_-}+ n_z\,\frac{\partial
\hat{H}^{L(R)}}{\partial \hat{k}_z}\right).
\end{equation}
Here $\hat{H}^L$ ($\hat{H}^R$) denotes the Hamiltonian, in which
the right(left)-hand operators $\hat{k}_+$, $\hat{k}_-$, and
$\hat{k}_z$ are treated as c-numbers, i.e. only the
left(right)-hand operators $\hat{k}_+$, $\hat{k}_-$, and
$\hat{k}_z$ are differentiated. Using the explicit form of the
Hamiltonian $\hat{H}$ (see Eq.~(\ref{H})), one finds that
$\hat{V}_r^L$ can be obtained multiplying by $1/\hbar$ the
Hamiltonian $\hat{H}$, in which all the left-hand operators
$\hat{k}_+$, $\hat{k}_-$, and $\hat{k}_z$ are replaced by $n_+$,
$n-$, and $n_z$, correspondingly, while all the terms that do not
contain the former operators are set to zero. It can be also shown
that
\begin{equation}\label{VelR}
\hat{V}_r^R(\xi,\chi)=\hat{\tau}\,\hat{V}_r^L(-\xi,-\chi),
\end{equation}
where the operator $\hat{\tau}$ draws all effective-mass
parameters through the operators $\hat{k}_+$, $\hat{k}_-$, and
$\hat{k}_z$ to the utmost right-hand positions.

The radial velocity $\hat{{\mathcal V}}_j^{(p),L}$ is obtained
from $\hat{V}_r^L$ in the same way as the radial Hamiltonian
$\hat{{\mathcal H}}_j^{(p)}$ was obtained from $\hat{H}$, i.e. by
the definition~(\ref{Hint}). For $n_+$, $n-$, and $n_z$ (see
Eq.~(\ref{n})) the expressions similar to (\ref{kY}) are valid if
one replaces $-i\,A_l^{(p)}$ by $-p$. Therefore, $\hat{{\mathcal
V}}_j^{(p),L}$ can be found by multiplying the Hamiltonian
$\hat{{\mathcal H}}_j^{(p)}$ by $i/\hbar$, replacing all the
left-hand operators $A_l^{(p)}$ by $-p$, and setting all the
terms, which do not contain the operator $A_l^{(p)}$, to zero.
This procedure results in
\begin{equation}\label{VelL}
\hat{{\mathcal V}}_j^{(p),L}=\frac{1}{2}\hat{{\mathcal
J}}_j^{(p)},
\end{equation}
where $\hat{{\mathcal J}}_j^{(p)}$ has been defined by
Eq.~(\ref{BC}). Using Eqs.~(\ref{Vel}), (\ref{VelL}), and
(\ref{VelR}) one obtains
\begin{equation}\label{rvelocity}
\hat{{\mathcal V}}_j^{(p)}=\frac{1}{2}\left(\hat{{\mathcal
J}}_j^{(p)}(\xi,\chi) +\hat{\tau}\,\hat{{\mathcal
J}}_j^{(p)}(-\xi,-\chi)\right).
\end{equation}
Considering the explicit form of the matrix $\hat{{\mathcal
J}}_j^{(p)}$ we see that all the terms, containing parameters $\xi$ and
$\chi$, responsible for the nonsymmetrical form of the Hamiltonian, cancel.
Consequently, we obtain
\begin{equation}\label{Velosity}
\hat{{\mathcal V}}_j^{(p)}=
\frac{1+\hat{\tau}}{2}\,\hat{{\mathcal J}}_j^{(p)}(\xi=0,\chi=0).
\end{equation}
If all the effective-mass parameters are piecewise-constant functions
of $r$, then at any point of the heterostructure except the spherical
heterointerfaces, we can replace $\hat{\tau}$ by $1$, and
therefore
\begin{equation}\label{VelosityH}
\hat{{\mathcal V}}_j^{(p)}=\hat{{\mathcal
J}}_j^{(p)}(\xi=0,\chi=0).
\end{equation}
It is clearly seen now, that the commonly used CRs for spherical
QDHs\cite{Sercel,Jaskolski} are the same as BCs (\ref{new}),
(\ref{BC}) obtained from the symmetrized Hamiltonian ($\xi=0$,
$\chi=0$). Like $\hat{{\mathcal J}}_j^{(p)}$, the radial
component of the velocity operator (\ref{Velosity}) is not
hermitian when $\hat{\tau}=1$. Therefore, in order to prove that both
current and velocity are conserved simultaneously, we should
verify whether the real parts of the current density and of the
velocity density are the same, i.e. we should check whether the
equality
\begin{equation}\label{eq}
\textrm{Re}\left[\left(R_j^{(p)}\right)^\dag\hat{{\mathcal
J}}_j^{(p)} R_j^{(p)}\right]=
\textrm{Re}\left[\left(R_j^{(p)}\right)^\dag\hat{{\mathcal
V}}_j^{(p)} R_j^{(p)}\right]
\end{equation}
holds true. Here $R_j^{(p)}$ is the radial wave function defined by
Eq.~(\ref{SchrE}). Substituting Eq.~(\ref{BC}) into the
left-hand-side part of Eq.~(\ref{eq}), we see that all the terms
containing the parameters $\xi$ and $\chi$ cancel, because their
contribution to the current density is purely imaginary.
Therefore, in conformity with Eq.~(\ref{VelosityH}), the equation
(\ref{eq}) is proven to be fair.

\section{Results of the calculation and discussion}

In this section we investigate the electronic structure of three
spherical QDHs with different values of the energy gaps: a
zero-gap semiconductor embedded into a wide-gap semiconductor
(HgS/CdS), a narrow-gap semiconductor embedded into a medium-gap
semiconductor (InAs/GaAs), and a medium-gap semiconductor
embedded into a wide-gap semiconductor (GaAs/AlAs). Note that in
these widely used experimentally relevant materials, the effective-mass
parameters are substantially different. The bulk 8-band
parameters of the used III-V and IV-VI materials are listed in Tables
\ref{GaAs} and \ref{HgS}, correspondingly. For electron and hole
levels, obtained within the spherical 8-band model, we use a
common notation: $nQ_j^{(e)}$ denotes an electron state and
$nQ_j^{(h)}$ denotes a hole state, where $n$ is the number of the
level with a given symmetry and $Q=S,P,D,\dots$ denotes the lowest
value of the momentum $l$ in the spherical harmonics of
Eq.~(\ref{wf2}) in front of the CB Bloch functions for an
electron state and in front of the VB Bloch functions for a hole
state, i.e. $Q=j-p/2$ for an electron and
$Q=\min(j+p/2,|j-3p/2|)$ for a hole.

\subsection{Electron energy levels}

\begin{figure}
\centering
\includegraphics[scale=1.5]{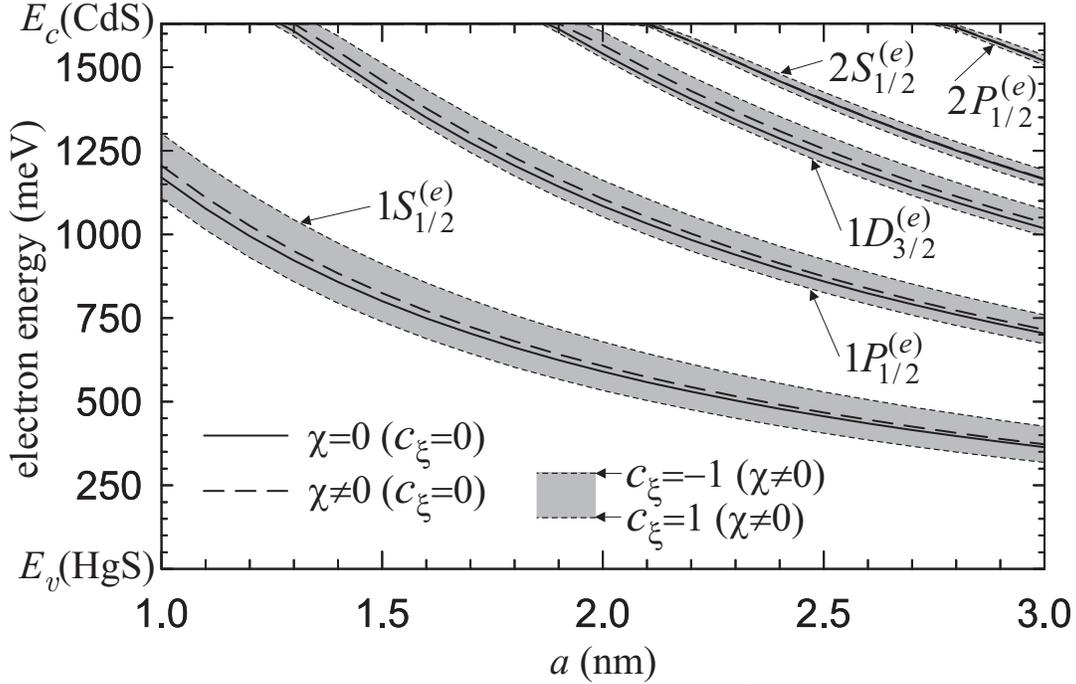}
\caption{All the discrete electron energy levels in the HgS/CdS
QDH as a function of the quantum dot radius. The $P_{3/2}^{(e)}$
and $D_{5/2}^{(e)}$ energy levels are not shown here and in
Figs.~\ref{fig:2} and \ref{fig:3} because in the chosen scale they
coincide with the levels $P_{1/2}^{(e)}$ and $D_{3/2}^{(e)}$,
correspondingly. Solid lines represent the result of the
symmetrized 8-band model ($c_{\xi}=0$, $\chi=0$). With the
nonsymmetrized valence band part of the Hamiltonian ($\chi\ne 0$),
dashed lines show the case $c_\xi=0$ while gray bands represent
the continuous change of $c_\xi$ from $1$ to $-1$. The gray bands
refer to a possible variation in energy due to conduction
band/valence band coupling via the position dependence of the
interband momentum matrix element. Here and in Figs.~\ref{fig:2} -
\ref{fig:6}, the insert with inscriptions $c_{\xi}=-1$ and
$c_{\xi}=1$ shows to which values of $c_\xi$ the edges of the gray
bands are related.}\label{fig:1}
\end{figure}

\begin{figure}
\centering
\includegraphics[scale=1.5]{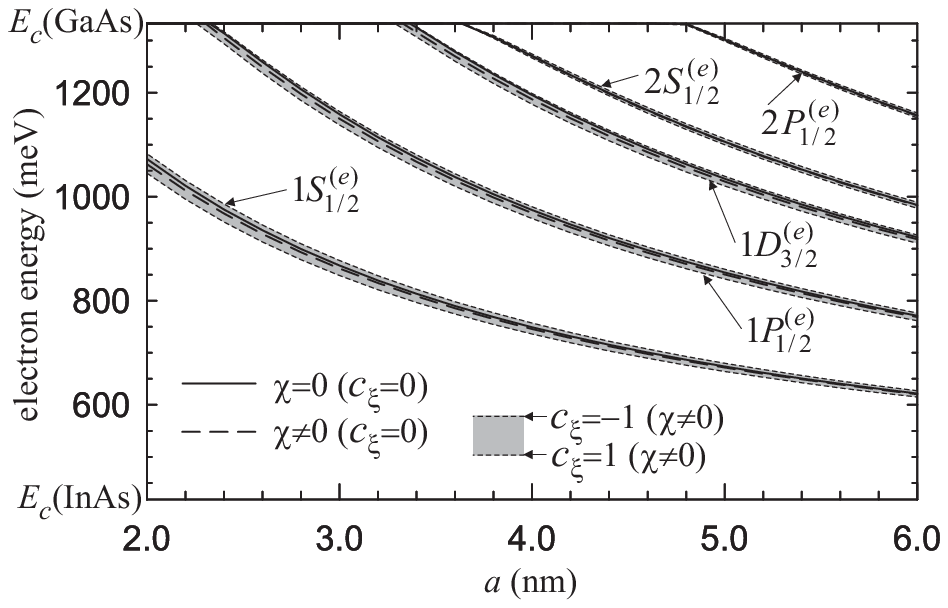}
\caption{All the discrete electron energy levels in the InAs/GaAs QDH
as a function of the quantum dot radius. Other denotations are the same as in
Fig.~\ref{fig:1}.}\label{fig:2}
\end{figure}

\begin{figure}
\centering
\includegraphics[scale=1.5]{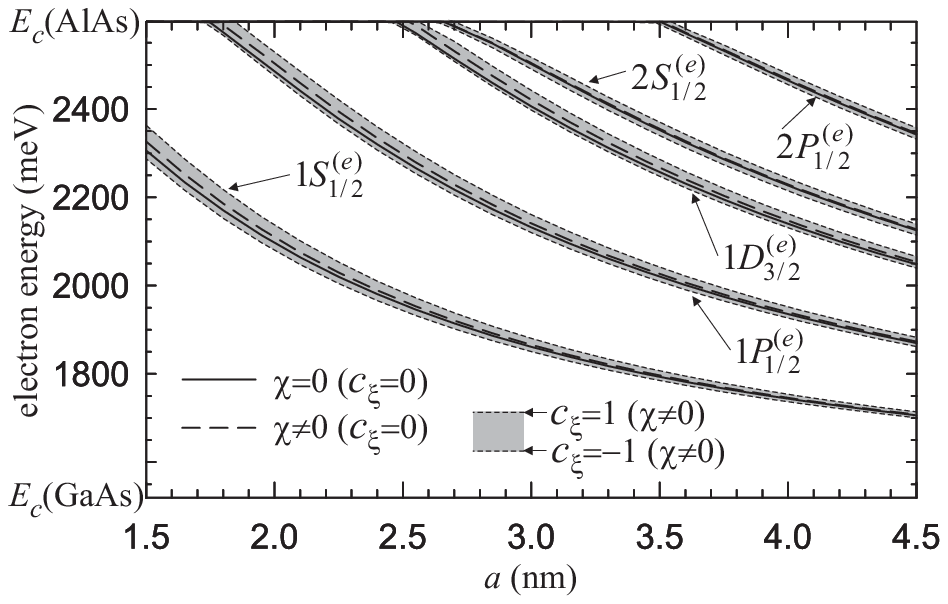}
\caption{All the discrete electron energy levels in the GaAs/AlAs QDH
as a function of the quantum dot radius. Other denotations are the same as in
Fig.~\ref{fig:1}.}\label{fig:3}
\end{figure}

The electron energy levels of the HgS/CdS, InAs/GaAs, and
GaAs/AlAs QDHs are depicted in Figs.~\ref{fig:1}-\ref{fig:3},
correspondingly, as a function of the quantum dot radius $a$. The
value of the spin-orbit splitting of electron energy levels is
small ($\approx$ 3~meV) for all considered QDHs (see
Table~\ref{table} and Appendix~B). Therefore, only the lowest
level of the pair of split levels is shown in
Figs.~\ref{fig:1}-\ref{fig:3}. For all examined QDHs, the lowest
level of such a pair is the level with the least total momentum
$j$.

Analyzing Figs.~\ref{fig:1}-\ref{fig:3} we arrive at the
following empirical formula, which determines the energy shift of
all electron levels with $n=1$ when a nonzero value of the
parameter $\chi$ is considered:
\begin{equation}\label{ee0}
E_e-\left.E_e\right|_{\chi=0}=-\frac{\hbar^2}{m_0
a^2}(\chi_1-\chi_2).
\end{equation}
Here the indices ``1'' and ``2'' denote interior and exterior
materials, correspondingly. For $n>1$ the value of this shift is
much smaller, than for $n=1$. It is seen that the shift
$E_e-\left.E_e\right|_{\chi=0}$ is about 16~meV for HgS/CdS ($a=$
2 nm) and about 4 meV for InAs/GaAs ($a=$ 4 nm) and GaAs/AlAs
($a=$ 3 nm) QDHs (see Table~\ref{table}). As provided by
Eq.~(\ref{ee0}), the value of this shift is inversely proportional
to the square of the quantum dot radius. Consequently, for large
QDHs one can use with high accuracy the symmetrized with respect
to $\chi$ Hamiltonian, and the expression~(\ref{ee0}) is the
measure of accuracy. If $\chi_1<\chi_2$, then the nonsymmetrized
energy level lies higher than the symmetrized one (see
Figs.~\ref{fig:1}, \ref{fig:3}), and if $\chi_1>\chi_2$, then the
nonsymmetrized energy level lies lower than the symmetrized one
(see Fig.~\ref{fig:2}).

Gray bands in Figs.~\ref{fig:1}-\ref{fig:3} reflect the change of
the parameter $c_\xi$ (see Eq.~(\ref{cxi})) from 1 to $-$1. The
chosen interval includes the following specific values of $c_\xi$:
$c_\xi=0$ (symmetrized Hamiltonian) and $c_\xi=\pm 1$ (see
Appendixes~A and B). With such a change of $c_\xi$, the electron
energy increases in Figs.~\ref{fig:1}, \ref{fig:2} and decreases
in Fig.~\ref{fig:3}. Therefore, the shift of an energy level with
$n=1$ with respect to the level position when $c_\xi=0$ can be
estimated by the formula:
\begin{equation}\label{eec}
E_e-\left.E_e\right|_{c_\xi=0}=b\,c_\xi\left(\sqrt{E_{p,1}}-
\sqrt{E_{p,2}}\right)\quad (b>0),
\end{equation}
where we take into account Eq.~(\ref{cxi}) and the fact that
$v_{1,2}\sim\sqrt{E_{p,1,2}}$. $E_p$ is the Kane energy (see
Tables~\ref{GaAs} and \ref{HgS}). For $n>1$ the shift
$E_e-\left.E_e\right|_{c_\xi=0}$ becomes much smaller, than for $n=1$.
The parameter $b$ in Eq.~(\ref{eec}) decreases with increasing $a$
and with increasing the energy gap in the
interior material. Such a behaviour of the parameter $b$ is
connected with the fact that it is proportional to
the value of hole radial components of the electron wave function
at the heterointerface. It is clear now that the observed strong
dependence of the energy levels in the HgS/CdS QDH on
$c_\xi$ (see Fig.~\ref{fig:1}) is due to the large value of the difference
$\sqrt{E_{p,1}}- \sqrt{E_{p,2}}$ and to the zero
energy gap in HgS. For two other QDHs the dependence of the energy levels
on $c_\xi$ is a few times weaker, than that for the HgS/CdS QDH.

Hence, Eqs.~(\ref{ee0}) and (\ref{eec}) allow us to estimate
corrections to the eigenenergies due to the replacement of the
heuristic symmetrized Hamiltonian with the nonsymmetrized
Hamiltonian avoiding complicated calculations. It follows from
Eq.~(\ref{ee0}) that such corrections rise with decreasing the
quantum dot radius as $1/a^2$. Therefore, one should use the
nonsymmetrized Hamiltonian for description of quantum dots with
small radii.

\subsection{Hole energy levels}

\begin{figure}
\centering
\includegraphics[scale=1]{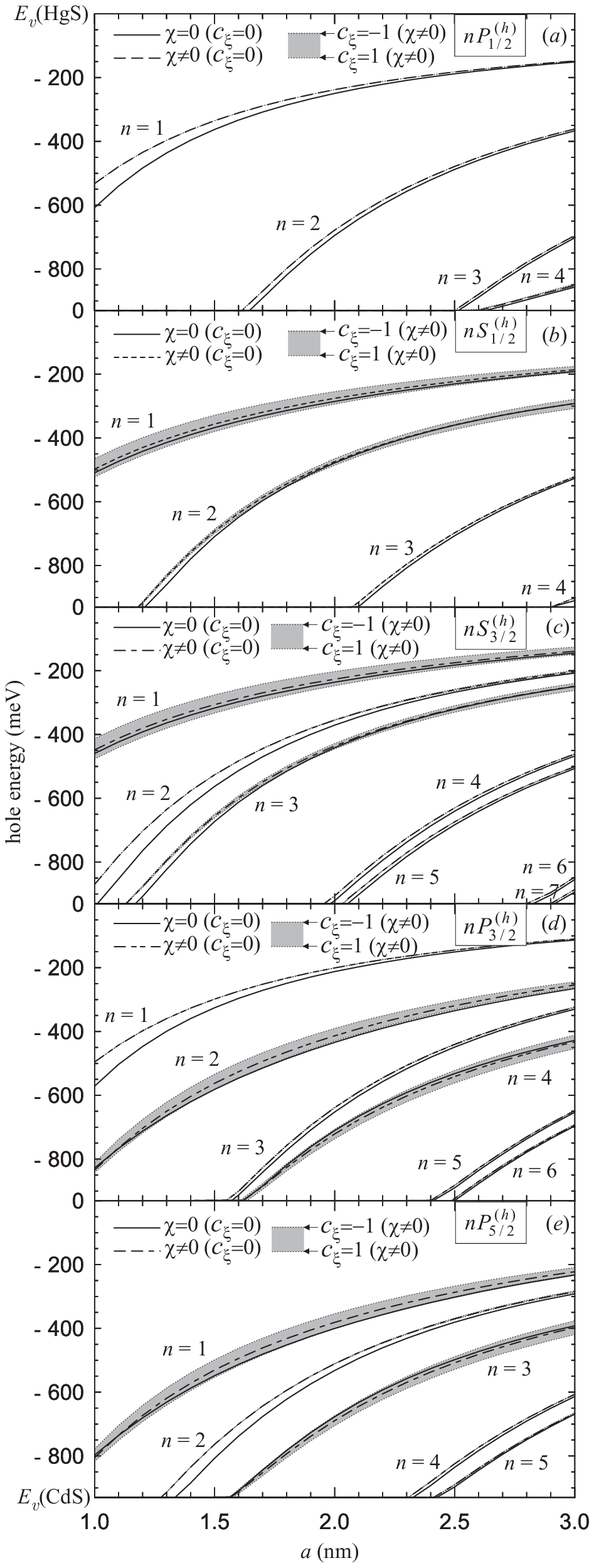}
\enlargethispage{100pt} \caption{All the discrete hole energy levels
of $S$- and $P$-types in the HgS/CdS QDH as a function of the quantum dot
radius. Other denotations are the same as in
Fig.~\ref{fig:1}.}\label{fig:4}
\end{figure}

\begin{figure}
\centering
\includegraphics[scale=1.5]{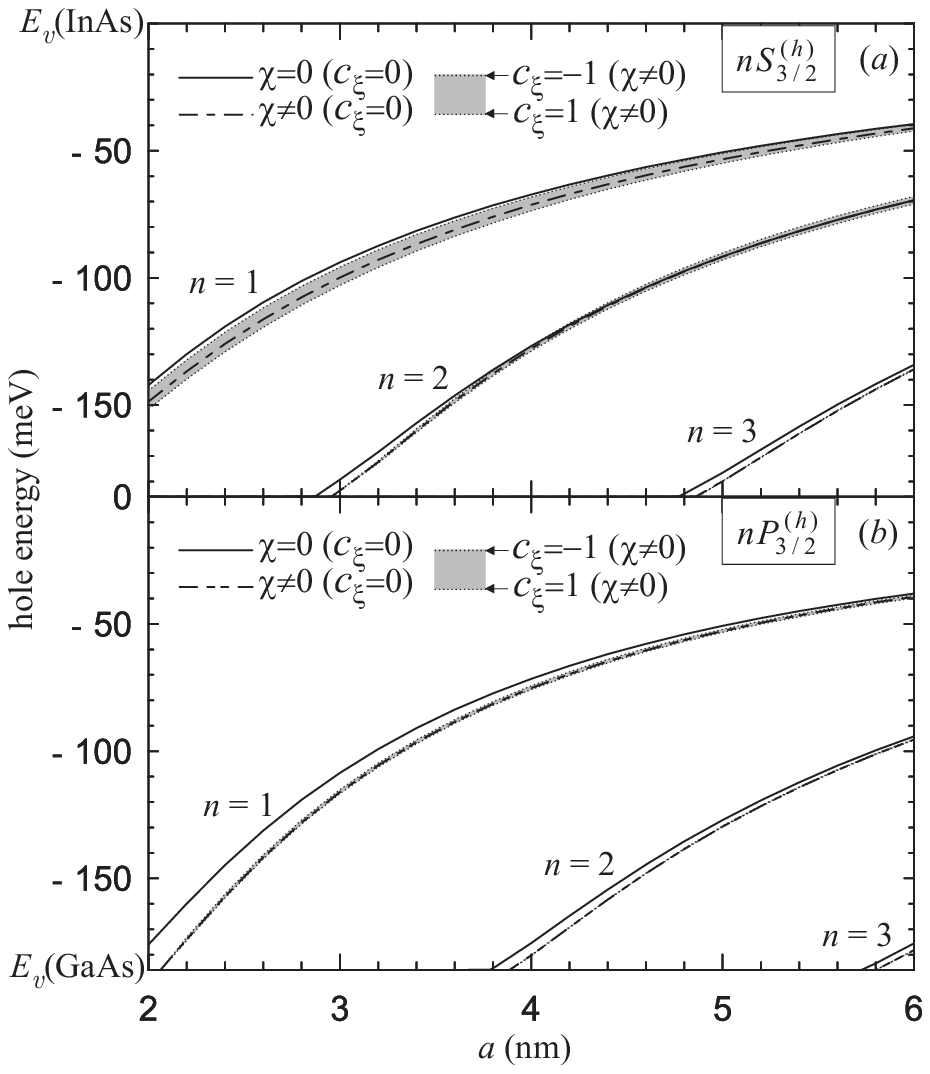}
\caption{All the discrete hole energy levels with $j=3/2$ in the
InAs/GaAs QDH as a function of the quantum dot radius. Other denotations
are the same as in Fig.~\ref{fig:1}.}\label{fig:5}
\end{figure}

\begin{figure}
\centering
\includegraphics[scale=1.5]{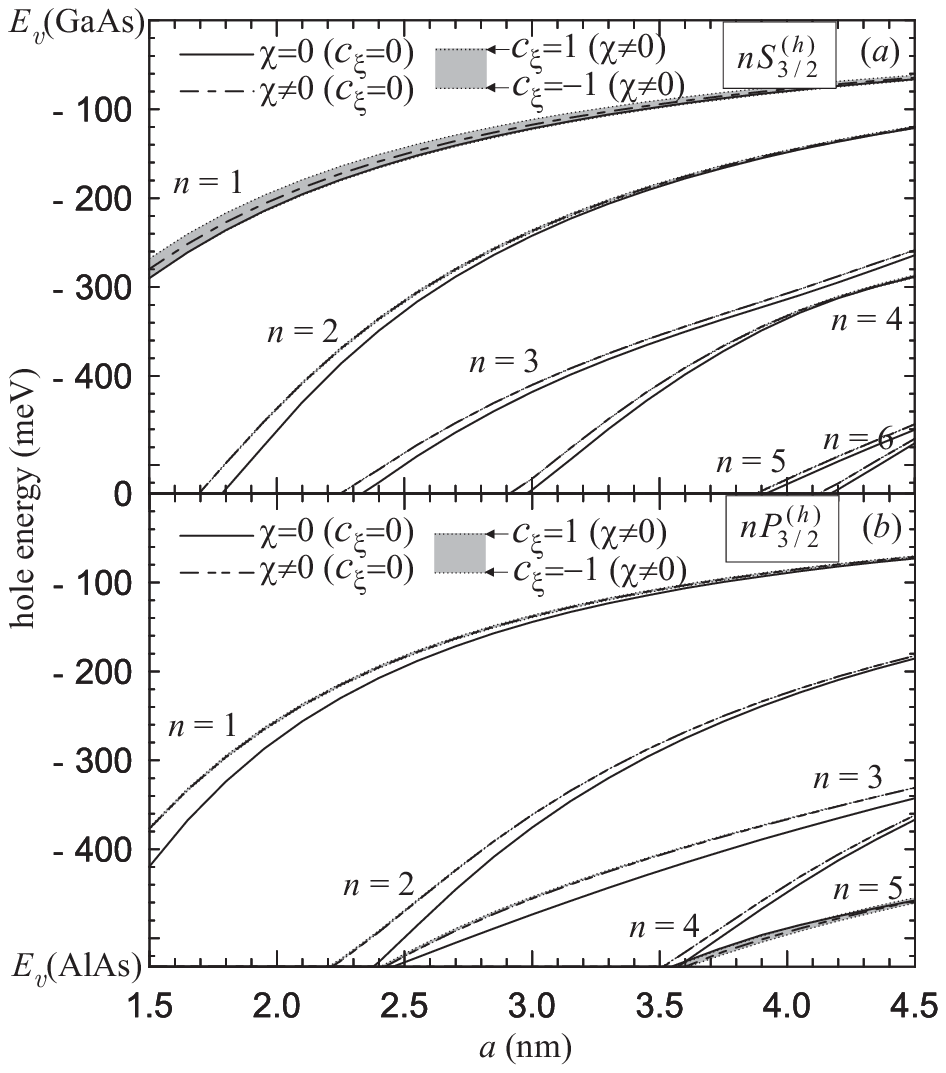}
\caption{All the discrete hole energy levels with $j=3/2$ in the
GaAs/AlAs QDH as a function of the quantum dot radius. Other denotations
are the same as in Fig.~\ref{fig:1}.}\label{fig:6}
\end{figure}

All the hole energy levels of $S$- and $P$-types in the HgS/CdS
QDH, with $j=3/2$ in the InAs/GaAs QDHs and with $j=3/2$ in the
GaAs/AlAs QDHs are depicted in Figs.~\ref{fig:4}, \ref{fig:5} and
\ref{fig:6}, correspondingly, as a function of the quantum dot
radius. It is seen from these figures that the empirical formula
(\ref{ee0}) holds for hole levels, too, i.e. for a nonzero value
of $\chi$, the hole energy levels shift in the same direction as
the electron energy levels do. For the HgS/CdS QDH ($a=$ 2 nm) the
shift of the hole ground state level is about 9 meV, what is
smaller than the shift of the electron ground state level. At the
same time, for InAs/GaAs ($a=$~4~nm) and GaAs/AlAs ($a=$~3~nm)
QDHs the shift for the hole ground state level is almost the same
as that for the electron ground state level (see
Table~\ref{table}). For the higher hole levels ($n>1$) the value
of the shift under consideration decreases with increasing $n$
much weaker than it does for the electron levels.

The dependence of the hole levels on the parameter $c_\xi$ is substantially
different from such a dependence for the electron levels. The
formula (\ref{eec}) can be approximately applied here only for
the level $1S_{3/2}^{(h)}$, which is the hole ground state energy
for all examined QDHs. It is seen that this energy level strongly
depends on $c_\xi$ even for InAs/GaAs and GaAs/AlAs QDHs. All the
other hole energy levels under analysis depend on $c_\xi$ very weakly,
and such a dependence is revealed only in Fig.~\ref{fig:4} for the HgS/CdS
QDH.

\subsection{Electron and hole wave functions and pair energies}

\begin{figure}
\centering
\includegraphics[scale=1.5]{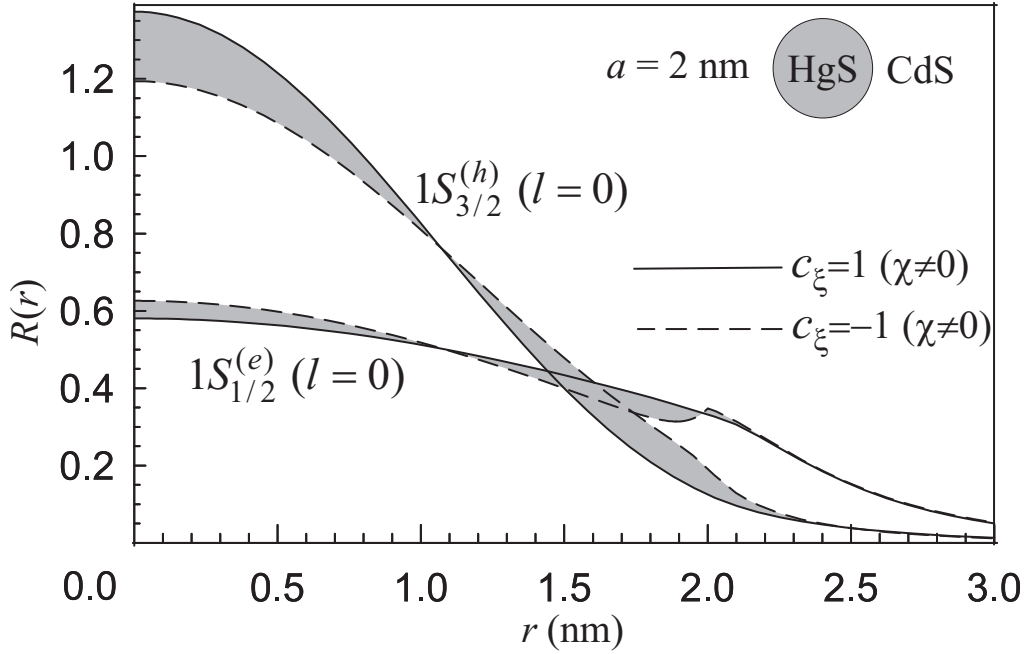}
\caption{$S$-type radial components of the wave functions of the
electron and hole ground states in the HgS/CdS QDH (radius $a$ =
2~nm) within the nonsymmetrized 8-band model. Solid and dashed
lines denote the cases $c_\xi=1$ and $c_\xi=-1$, correspondingly,
while gray bands represent the continuous change of $c_\xi$ within
these limits. Each radial wave function is normalized by unity,
i.e. the integral probability $\sum_\mu \int_0^\infty r^2 R^2_\mu
(r) dr=1$, where $\mu$ labels the radial components. Contributions
to the integral probability from the depicted radial components
vary from 68.6~\% to 64.8~\% for an electron and from 74.3~\% to
81.1~\% for a hole when $c_\xi$ changes from $1$ to $-1$. At the
same time the electron energy changes from 533.6~meV to 678.8~meV
and the hole energy changes from $-$247.2~meV to
$-$202.1~meV.}\label{fig:7}
\end{figure}

\begin{figure}
\centering
\includegraphics[scale=1.5]{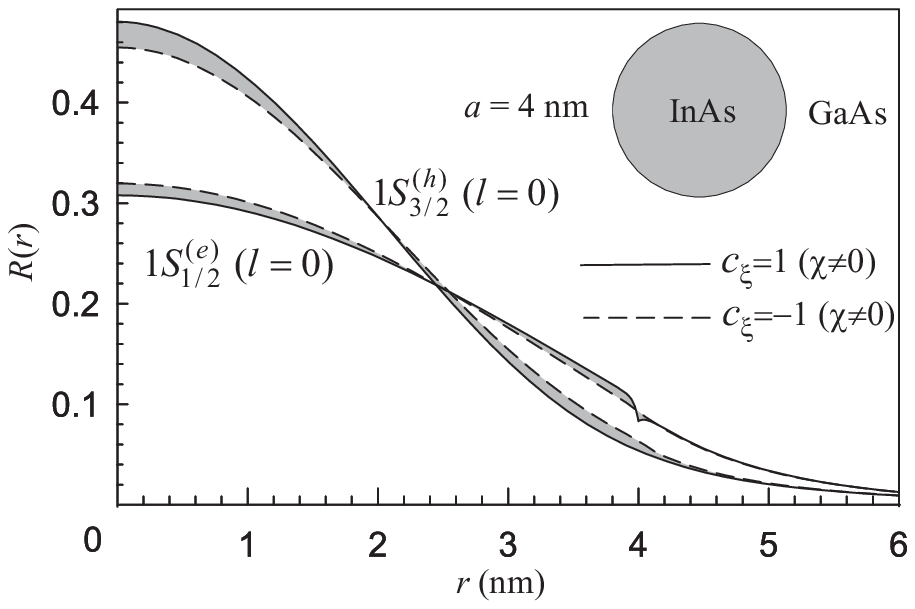}
\caption{$S$-type radial components of the wave functions of the
electron and hole ground states in the InAs/GaAs QDH (radius $a$ =
4~nm) within the nonsymmetrized 8-band model. Normalization of
each radial wave function and denotations are same as in
Fig.~\ref{fig:7}. Contributions to the integral
probability from the depicted radial components vary from 80.6~\%
to 79.2~\% for an electron and from 75.3~\% to 78.7~\% for a hole
when $c_\xi$ changes from $1$ to $-1$. At the same time the
electron energy changes from 736.0~meV to 757.1~meV and the hole
energy changes from $-$73.6~meV to $-$68.2~meV.}\label{fig:8}
\end{figure}

\begin{figure}
\centering
\includegraphics[scale=1.5]{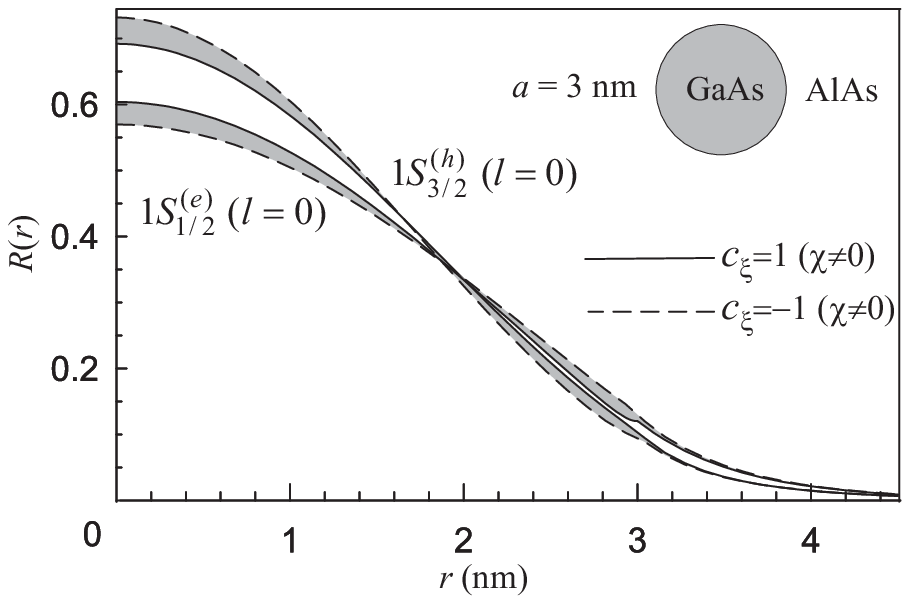}
\caption{$S$-type radial components of the wave functions of the
electron and hole ground states in the GaAs/AlAs QDH (radius $a$ =
3~nm) within the nonsymmetrized 8-band model. Normalization of
each radial wave function and denotations are the same as in
Fig.~\ref{fig:7}. Contributions to the integral
probability from the depicted radial components vary from 85.4~\%
to 86.9~\% for an electron and from 88.9~\% to 86.6~\% for a hole
when $c_\xi$ changes from $1$ to $-1$. At the same time the
electron energy changes from 1880.4~meV to 1850.4~meV and the hole
energy changes from $-$111.6~meV to $-$122.9~meV.}\label{fig:9}
\end{figure}

In Figs.~\ref{fig:7}, \ref{fig:8}, and \ref{fig:9}, the $S$-components of
the radial wave functions of the electron ground state
($1S_{1/2}^{(e)}$) and of the hole ground state
($1S_{3/2}^{(h)}$) are depicted for HgS/CdS ($a=$ 2 nm), InAs/GaAs
($a=$ 4 nm), and GaAs/AlAs ($a=$ 3 nm) QDHs, correspondingly. It
is seen that in all these QDHs, the hole density in the interior
material is higher than the electron density, and the electron
density is higher in the exterior material. It is also seen that
when $c_\xi$ changes from 1 to $-$1, the electron density in the
centers of the HgS/CdS and InAs/GaAs QDHs increases and the hole
density decreases. The opposite trends of behavior of the electron and hole
densities are observed in the center of the GaAs/AlAs QDH. The
abrupt change of the derivative of the electron radial component
with the change of $c_\xi$ is well seen at the heterointerfaces
of all QDHs under consideration. At the same time, the derivative
of the hole radial component changes smoothly. The contribution
of the hole radial components to the density of the electron
state (at $c_\xi$=0) is as high as 33~\% for HgS/CdS, 20~\% for InAs/GaAs, and
14~\% for GaAs/AlAs QDH. Such contributions show that the
nonparabolicity of the electron dispersion law is substantial
even for the QDs of the medium-gap semiconductors (GaAs) and
certainly should be taken into consideration when the QDs of the narrow-gap
semiconductors (InAs) are investigated. The contribution of the
electron radial component to the density of the hole state (at $c_\xi$=0) is
6~\% for HgS/CdS, 1~\% for InAs/GaAs, and 1~\% for
GaAs/AlAs QDH. This fact leads to the conclusion that the additional
nonparabolicity of the hole dispersion law connected with the
influence of the conduction band can be neglected for both narrow-
and medium-gap semiconductor QDs.

\begin{figure}
\centering
\includegraphics[scale=1.5]{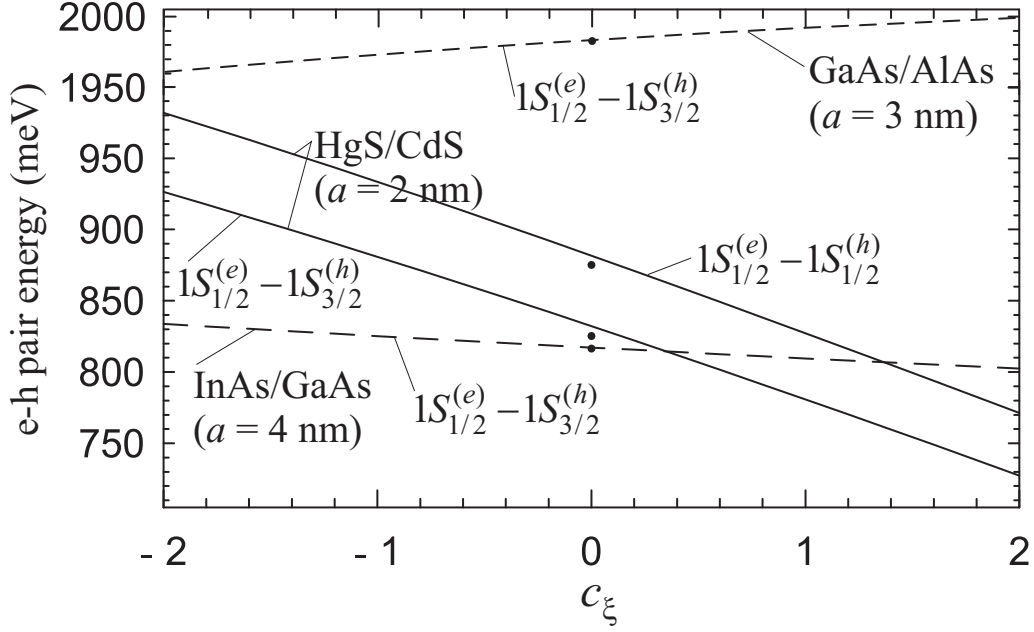}
\caption{The lowest electron-hole pair energies in different QDHs
as a function of $c_\xi$ ($\chi\ne 0$). Each dot is related to the
closest curve and indicates the corresponding result of the
symmetrized model ($c_{\xi}=0$, $\chi=0$).}\label{fig:10}
\end{figure}

Taking into account the principal role of the dissymmetry
coefficient $c_\xi$, one can evaluate the influence of this
parameter on the observable effects. With this purpose we
calculate the lowest electron-hole pair energies as a function of
$c_\xi$ for all QDHs under consideration (see Fig.~\ref{fig:10}). It is
seen from Fig.~\ref{fig:10} that when the parameter $c_\xi$
changes from $-$2 to 2, the corresponding energy differences
$E_{e-h}(c_\xi=2)-E_{e-h}(c_\xi=-2)$ constitute $-$175~meV for HgS/CdS
($a=$ 2 nm), $-$15~meV for InAs/GaAs ($a=$ 4 nm), and 20~meV for
GaAs/AlAs ($a=$ 3 nm) QDHs. These differences should be quite
accessible for the experimental detection.

\section{Conclusions}

The exact nonsymmetrized 8-band effective-mass Hamiltonian for an
arbitrary 3-dimensional heterostructure has been obtained using
the Burt's envelope-function representation. The 2$\times$2 electron
and 6$\times$6 hole energy-dependent Hamiltonians have been
deduced. Within the spherical approximation, the 8$\times$8,
2$\times$2, and 6$\times$6 radial Hamiltonians and the necessary BCs
have been derived for spherical QDHs. The boundary conditions for
radial symmetrized and nonsymmetrized Hamiltonians are different
and lead therefore to different energy levels and wave functions. We
have shown, further, that the CRs, which are commonly used to match the
solutions of the appropriate bulk
\textbf{\textit{k}}$\cdot$\textbf{\textit{p}} Hamiltonians,
coincide with BCs for the symmetrized Hamiltonians. A theoretical
estimate for the value of the spin-orbit splitting of electron
levels has been found. The energy levels of the nonsymmetrized 8-band
Hamiltonian have been calculated as a function of the dot radius
for three spherical QDHs: a zero-gap semiconductor embedded into
a wide-gap semiconductor (HgS/CdS), a narrow-gap semiconductor
embedded into a medium-gap semiconductor (InAs/GaAs), and a
medium-gap semiconductor embedded into a wide-gap semiconductor
(GaAs/AlAs). It has been demonstrated that parameters of
dissymmetry $\chi({\bf r})$ and $\xi({\bf r})$, giving nonzero
contribution to the multiband Hamiltonians only at the
heterointerfaces, have, nevertheless, a strong effect on
the electron and hole spectra. Thus, for practically important cases
of relatively small QDHs with
noticeably different effective-mass parameters of the constituent
materials, the use of the obtained Hamiltonian is necessary for the adequate
description of experiment.

\section*{Acknowledgements}

This work has been supported by the GOA BOF UA 2000, IUAP, FWO-V
projects G.0287.95, G.0274.01N and the W.O.G. WO.025.99N (Belgium).

\appendix

\section{Interband momentum matrix elements}

Using the general effective mass equations [Eq.~(6.3) of
Ref.~\onlinecite{Burt1}], one can see the origin of the interband
momentum matrix elements $v_1({\bf r})$ and $v_2({\bf r})$ from
Eq.~(\ref{H44}):
\begin{equation}\label{v1d}
v_1(E,{\bf r})=-\frac{4i}{\hbar} \langle S|\hat{p}_z|Z\rangle
-\frac{4i}{\hbar} \sum\limits_{\nu} (E-H_{\nu\nu}({\bf r}))^{-1}
H_{S\nu}({\bf r})\, \langle \nu|\hat{p}_z|Z\rangle;
\end{equation}
\begin{equation}\label{v2d}
v_2(E,{\bf r})= -\frac{4i}{\hbar} \sum\limits_{\nu}
(E-H_{\nu\nu}({\bf r}))^{-1} \langle S|\hat{p}_z|\nu\rangle\,
H_{\nu Z}({\bf r}).
\end{equation}
To obtain Eqs.~(\ref{v1d}) and (\ref{v2d}) it should be taken into
account that within the developed 8-band approach the conduction
band with the Bloch function $|S\rangle$ and the valence band with
the Bloch functions $|X\rangle$, $|Y\rangle$, and $|Z\rangle$ are
included explicitly, while all other bands with the Bloch
functions $|\nu\rangle$ are considered to be remote. Further,
following the technique of Ref.~\onlinecite{Burt1} it is necessary
to exclude the energy dependence from Eqs.~(\ref{v1d}) and
(\ref{v2d}) by replacing $E$ with an average energy, for instance
the energy at the middle of the narrowest gap for the
heterostructure compounds. Parameters $v_1({\bf r})$ and $v_2({\bf
r})$ are approximately considered to be constant in each layer of
a heterostructure.

In bulk, if Burt's material-independent basis functions coincide
with the bulk Bloch functions, the second term in the r.h.s. of
Eq.~(\ref{v1d}) and the r.h.s. of Eq.~(\ref{v2d}) vanish, because
in this case $H_{S\nu}({\bf r})=0$ and $H_{\nu Z}({\bf r})=0$.
Therefore, one obtains $v_1=2v$ ($v=-2i\langle
S|\hat{p}_z|Z\rangle/\hbar$) and $v_2=0$, what results in $\xi=v$
(see Eq.~(\ref{xi})) and $c_{\xi}=1$ (see Eq.~(\ref{cxi})). When
materials constituting the heterostructure have close parameters,
the second term in the r.h.s. of Eq.~(\ref{v1d}) and the r.h.s. of
Eq.~(\ref{v2d}) are small compared with the first term in the
r.h.s. of Eq.~(\ref{v1d}). In this case $\xi$ does not differ
significantly from $v$, and $c_{\xi}$ is close to 1. In a general
case of disparate materials, $c_{\xi}$ can take arbitrary values.

\section{Energy-dependent separate Hamiltonians for electrons and
holes}

For narrow-gap semiconductors, the accurate way to take into
account the coupling of conduction and valence bands, is to
consider the 8-band Hamiltonian. However, sometimes it is easier
to solve a CB or VB Schr\"odinger equation with energy-dependent
effective-mass parameters. Solutions of these equations are just
an approximation to the results of the 8-band model. In what
follows, we deduce the 2$\times$2 energy-dependent Hamiltonian
for an electron and the 6$\times$6 energy-dependent Hamiltonian for a
hole from the exact 8-band nonsymmetrized effective mass
Hamiltonian.

\subsection{2$\times$2 energy-dependent Hamiltonian for an electron}

\subsubsection{Nonsymmetrized CB Hamiltonian}

We start with the nonsymmetrized 8-band Hamiltonian $\hat{H}$
defined by Eq.~(\ref{H}). The wave function $\Psi$, i.e. a vector
of eight envelope functions $\Psi_1$, \dots, $\Psi_8$ is an
eigenfunction of the matrix Schr\"odinger equation
\begin{equation}\label{SE}
\hat{H}\,\Psi=E\,\Psi,
\end{equation}
where $E$ is an eigenenergy. To find the CB Hamiltonian, one
should treat all VBs as remote. Therefore, we should exclude all
VB envelopes, i.e. $\Psi_3$, \dots, $\Psi_8$, from
Eq.~(\ref{SE}). As seen, this exclusion is possible only within
the approximation $\gamma_1=0$, $\gamma_2=0$, and $\gamma_3=0$, in
other words, when the contributions to the hole effective-mass
parameters from the remote bands (all bands except two CBs and six
VBs) is negligible. This is a very close approximation, because
the parameters $\gamma_1$, $\gamma_2$, and $\gamma_3$ are small
for almost all materials and, determining contributions to the VB, they
certainly have small influence on the electron levels. Under this
approximation $\chi=-1/3$ (see Eq.~(\ref{chidef})) and it cancels
from the Hamiltonian. Another necessary approximation is
$c_\xi=-1$, i.e. $\xi=-v$, and therefore $v_1=0$, $v_2=2v$ (see
Eqs.~(\ref{cxi}) and (\ref{xi})). This is the only approximation
that does not lead to the discontinuity of CB envelopes $\Psi_1$
and $\Psi_2$ at the heterointerface. Now we can express six VB
envelopes $\Psi_3$, \dots, $\Psi_8$ in terms of two CB envelopes
$\Psi_1$ and $\Psi_2$ using the last six equations of the
set~(\ref{SE}). This procedure results in
\begin{equation}\label{cb-vb}
\begin{array}{rcl}
\Psi_3&=&-i\displaystyle\frac{v}{\varepsilon-\varepsilon_v}\hat{k}_-
\Psi_1,\\[12pt]
\Psi_4&=&\displaystyle\frac{v}{\varepsilon-\varepsilon_v}
\left(\sqrt{\displaystyle\frac{2}{3}}\,\hat{k}_z\Psi_1-
\displaystyle\frac{1}{\sqrt 3}\,\hat{k}_-\Psi_2\right),\\[12pt]
\Psi_5&=&\displaystyle\frac{v}{\varepsilon-\varepsilon_v}
\left(\displaystyle\frac{-i}{\sqrt 3}\,\hat{k}_+\Psi_1-i
\sqrt{\displaystyle\frac{2}{3}}\,\hat{k}_z\Psi_2\right),\\[12pt]
\Psi_6&=&-\displaystyle\frac{v}{\varepsilon-\varepsilon_v}\hat{k}_+
\Psi_2,\\[12pt]
\Psi_7&=&\displaystyle\frac{v}{\varepsilon-\varepsilon_v+\delta}
\left(\displaystyle\frac{-i}{\sqrt 3}\,\hat{k}_z\Psi_1-i
\sqrt{\displaystyle\frac{2}{3}}\,\hat{k}_-\Psi_2\right),\\[12pt]
\Psi_8&=&\displaystyle\frac{v}{\varepsilon-\varepsilon_v+\delta}
\left(\sqrt{\displaystyle\frac{2}{3}}\,\hat{k}_+\Psi_1-
\displaystyle\frac{1}{\sqrt 3}\,\hat{k}_z\Psi_2\right),
\end{array}
\end{equation}
where $\varepsilon=2m_0E/\hbar^2$. Substituting the envelopes
(\ref{cb-vb}) into the first two equations of the set~(\ref{SE}),
one obtains the sought CB Hamiltonian for the electron envelopes
$\Psi_1$ and $\Psi_2$. This Hamiltonian has the form
\begin{equation}\label{22H}
\hat{H}_e=\frac{\hbar^2}{2m_0}\left(
\begin{array}{cc}
\varepsilon_c+P_e & C_e\\
C_e^\dag & \varepsilon_c+P_e^*
\end{array}
\right),
\end{equation}
where
$$
P_e=\hat{k}_+\left(\frac{m_0}{m_c(\varepsilon)}
+g_c(\varepsilon)\right)\hat{k}_-+\hat{k}_-\left(\frac{m_0}{m_c(\varepsilon)}
-g_c(\varepsilon)\right)\hat{k}_+
+\hat{k}_z\frac{m_0}{m_c(\varepsilon)}\hat{k}_z,
$$
\begin{equation}\label{Ce}
C_e=-\sqrt{2}\left(\hat{k}_z\,g_c(\varepsilon)\,\hat{k}_-
-\hat{k}_-\,g_c(\varepsilon)\,\hat{k}_z\right),
\end{equation}
$$
\frac{m_0}{m_c(\varepsilon)}=\alpha+\frac{v^2}{3}
\left(\frac{2}{\varepsilon-\varepsilon_v}+
\frac{1}{\varepsilon-\varepsilon_v+\delta}\right),
$$
\begin{equation}\label{gc}
g_c(\varepsilon)=\frac{v^2}{3}
\left(\frac{1}{\varepsilon-\varepsilon_v}-
\frac{1}{\varepsilon-\varepsilon_v+\delta}\right).
\end{equation}
Here, $m_0/m_c(\varepsilon)$ is the inverse of the
energy-dependent effective mass of an electron, and
$g_c(\varepsilon)$ is an energy-dependent interfacial parameter,
which vanishes when the spin-orbit splitting $\delta$ is zero.
Like $\chi$, this parameter gives a nonzero contribution to the
Hamiltonian only at the heterointerface. The parameter
$g_c(\varepsilon)$ is responsible for the nonsymmetrical form of
the Hamiltonian (\ref{22H}) and for the mixing of the envelopes
$\Psi_1$ and $\Psi_2$. When one solves the Schr\"odinger equation
for an electron using the energy-dependent Hamiltonian
(\ref{22H}), one finds the eigenenergy $E$ and eigenfunctions
$\Psi_1$ and $\Psi_2$. Then, substituting these eigenfunctions
into Eq.~(\ref{cb-vb}), one obtains the rest envelopes $\Psi_3$,
\dots, $\Psi_8$, and therefore $\Psi$. While the functions
$\Psi_1$ and $\Psi_2$ are continuous at the heterointerface, the
functions $\Psi_3$, \dots, $\Psi_8$, as the envelope functions of
all the other remote bands, are not.\cite{Burt1} Note, that only
the wave functions $\Psi$, i.e. vectors of eight envelope
functions $\Psi_1$, \dots, $\Psi_8$ are orthonormalized. The
envelopes $\Psi_1$ and $\Psi_2$ are neither orthogonal nor
properly normalized. When the nonparabolicity is not strong, in
other words, when $m_c(\varepsilon)$ only weakly depends on the
energy, it is possible to choose one appropriate value of the
energy, e.~g. $\varepsilon_0$, to find $m_c(\varepsilon_0)$ and
$g_c(\varepsilon_0)$, and to substitute them into the Hamiltonian
(\ref{22H}). In such a way one obtains the Hamiltonian
$\hat{H}_e(\varepsilon_0)$. The eigenfunctions $\Psi_1$ and
$\Psi_2$ of this Hamiltonian will be orthonormalized and the rest
six envelopes will be no longer needed.

\subsubsection{CB Hamiltonian and BCs for a spherical QDH}

The radial CB Hamiltonian $\hat{{\mathcal H}}_{e,j}^{(p)}$ for
spherical QDHs can be derived from the Hamiltonian (\ref{22H}) by
the same way as the radial Hamiltonian (\ref{Hjp}) has been
obtained from the Hamiltonian (\ref{H}) in the spherical
approximation (see Sec.~III). Thus we find
\begin{equation}\label{elH}
\hat{{\mathcal H}}_{e,j}^{(p)}=\frac{\hbar^2}{2m_0}\left(
\varepsilon_c-{\mathcal P}_{e,j-p/2}^{(p)}\right),
\end{equation}
where
\begin{equation}\label{elP}
{\mathcal P}_{e,l}^{(p)}=\frac{(l+1)\Delta_l^{(1)}\left(
\displaystyle\frac{m_0}{m_c(\varepsilon)}+g_c(\varepsilon)\right)
+l\,\Delta_l^{(-1)}\left(\displaystyle\frac{m_0}{m_c(\varepsilon)}
+g_c(\varepsilon)\right)}{2l+1}-\Delta_l^{(p)}\Big(g_c(\varepsilon)\Big)
\end{equation}
and the operator $\Delta_l^{(p)}(\beta)$ is defined by
Eq.~(\ref{Delta}). Inside the $i$-th spherical layer, the
Hamiltonian (\ref{elH}) takes the form
\begin{equation}\label{elHi}
\hat{{\mathcal H}}_{e,j}^{i,(p)}=\frac{\hbar^2}{2m_0}\left(
\varepsilon_c-\frac{m_0}{m_{c,i}(\varepsilon)}\Delta_{j-p/2}\right),
\end{equation}
where $\Delta_l$ is the spherical Laplacian and
$m_{c,i}(\varepsilon)$ is the energy-dependent CB mass of the
$i$-th material. Further, one should solve the Schr\"odinger
equation with the Hamiltonian (\ref{elHi}) for each spherical
layer and match the obtained solutions at the spherical
heterointerfaces using the BCs (\ref{new}) (see Sec.~IV). The
radial component of the CB current operator $\hat{{\mathcal
J}}_{e,j}^{(p)}$ is obtained from the Hamiltonian (\ref{elH}) by
the same way as the radial component of the current operator
(\ref{BC}) has been obtained from the Hamiltonian (\ref{Hjp}).
Thus,
\begin{equation}\label{Je}
\hat{{\mathcal J}}_{e,j}^{(p)}=
\frac{i\hbar}{m_0}\left(-\frac{m_0}{m_c(\varepsilon)}
\frac{\partial}{\partial
r}-p(j+1/2-p)\frac{g_c(\varepsilon)}{r}\right).
\end{equation}

In a two-layer spherical QDH the electron energy depends on the
difference $g_{c,1}(\varepsilon)-g_{c,2}(\varepsilon)$ (as seen
from the BCs (\ref{new}), (\ref{Je})), where the indices ``1'' and
``2'' denote the interior and exterior materials, correspondingly.
The value of this difference is usually very small for typical QDHs.
Therefore, in the first approximation, one can find the energy
spectrum $E_l$ ($l=j-p/2$) neglecting the term proportional to
$p(j+1/2-p)\Big(g_{c,1}(\varepsilon)-g_{c,2}(\varepsilon)\Big)$
in the BCs. Then, including this term as a perturbation, one finds
the energy spectrum $E^j_l$. It is seen that the energy levels
with $l=0$ remain unchanged, while each energy level $E_l$ with
$l\ge 1$ splits into two levels: $E_l^{l+1/2}$ and $E_l^{l-1/2}$.
For the electron levels that are not very close to the CB minimum
(it is the case for the QDHs under consideration), the following estimate
can be obtained
\begin{equation}\label{estimate}
E_l=\frac{(l+1)\,E_l^{l+1/2}+l\,E_l^{l-1/2}}{2l+1},\quad
E_l^{l+1/2}-E_l^{l-1/2}=\frac{\hbar^2(2l-1)}{m_0 a^2}
\Big(g_{c,1}(\varepsilon_l)-g_{c,2}(\varepsilon_l)\Big),
\end{equation}
where $\varepsilon_l=2m_0E_l/\hbar^2$. In Table~\ref{table} we
have used Eq.~(\ref{estimate}) to estimate the spin-orbit
splitting of the lowest $P$ and $D$ levels. It is seen that the
value of the splitting of the electron levels is of the order of 3~meV
in all considered QDHs, and therefore this splitting can be
neglected. This fact does not imply that the dependence of the parameters of the
CB Hamiltonian on the energy can be neglected, too. As seen from
Table~\ref{table}, the electron effective masses in QDHs can
differ by a factor of 2 from their values in the corresponding
bulk materials.

\subsection{6$\times$6 energy-dependent Hamiltonian for a hole}

\subsubsection{Nonsymmetrized VB Hamiltonian}

The deduction of the VB Hamiltonian is analogous to the deduction
of the CB Hamiltonian with the only difference: one should treat
two CBs as remote. In order to express the CB envelopes $\Psi_1$
and $\Psi_2$ in terms of the VB envelopes $\Psi_3$, \dots,
$\Psi_8$ and to exclude them from the Schr\"odinger equation
(\ref{SE}), one should apply the approximation $\alpha=0$ and
$c_\xi=1$. This approximation has the same grounds as the
approximation used above to obtain the CB Hamiltonian. Now, we express
the CB envelopes $\Psi_1$ and $\Psi_2$ in terms of the VB
envelopes $\Psi_3$, \dots, $\Psi_8$ from the first two equations
of the set (\ref{SE}) and substitute them into the last six
equations of the same set. As a result we have the VB Hamiltonian
$\hat{H}_h$, which coincides with the Hamiltonian $\hat{H}$ (see
Eq.~(\ref{H})) where the first two rows and the first two columns are
deleted and the effective-mass parameters are changed in the
following way:
\begin{eqnarray}\label{parL}
\gamma_1&\rightarrow&\gamma_1^L(\varepsilon)=\gamma_1
+\frac{v^2}{3(\varepsilon_c-\varepsilon)},\nonumber\\
\gamma_{2,3}&\rightarrow&\gamma_{2,3}^L(\varepsilon)=\gamma_{2,3}
+\frac{v^2}{6(\varepsilon_c-\varepsilon)}.
\end{eqnarray}
Here, $\gamma_i^L(\varepsilon)$ are the energy-dependent Luttinger
parameters. In conformity with Eq.~(\ref{chidef}), one should
change the parameter of dissymmetry $\chi$ as follows
\begin{equation}\label{chiL}
\chi\rightarrow \chi^L(\varepsilon)=\chi
+\frac{v^2}{6(\varepsilon_c-\varepsilon)}.
\end{equation}
As a result of the change (\ref{chiL}), the parameter of dissymmetry
increases (see Table~\ref{table}), therefore the results
of the symmetrized Hamiltonian (with $\chi^L=0$) will deviate
sharply from the exact solutions. The parameters $\gamma_i^L$ usually
weakly depend on the energy. Consequently, to obtain the hole
spectrum one can use the Hamiltonian $\hat{H}_h(\varepsilon_0)$,
where $\varepsilon_0$ is an average hole energy.

\subsubsection{VB Hamiltonian and BCs for a spherical QDH}

The radial VB Hamiltonian $\hat{{\mathcal H}}_{h,j}^{(p)}$ for a
spherical QDH coincides with the radial Hamiltonian (\ref{Hjp})
in which $\gamma_1\rightarrow \gamma_1^L(\varepsilon)$ and
$\chi\rightarrow \chi^L(\varepsilon)$ (in conformity with
Eqs.~(\ref{parL}) and (\ref{chiL})), $\gamma\rightarrow
\gamma^L(\varepsilon)=\gamma
+\displaystyle\frac{v^2}{6(\varepsilon_c-\varepsilon)}$ (in
conformity with Eqs.~(\ref{parL}) and (\ref{gamma8})) and where
the first two rows and the first two columns are deleted. For the
radial components of the hole wave function one should use the
BCs~(\ref{new}), in which the radial component of the current
operator $\hat{{\mathcal J}}_{h,j}^{(p)}$ is given by the matrix
(\ref{BC}) where the first row and the first column are deleted
and the parameters $\gamma_1$, $\gamma$, $\chi$ are replaced by
the parameters $\gamma_1^L$, $\gamma^L$, $\chi^L$,
correspondingly.

% begin TABLES

\begin{table}
\caption{The 8-band effective-mass parameters of some III-V
materials. The VB offset $E_v$ is chosen to be zero in GaAs.
The parameters $\alpha$, $\gamma_1$, $\gamma$, and $\chi$ of the
spherical model are calculated from the listed here
effective-mass parameters.}\label{GaAs}
\begin{tabular}{lccc}
Parameters & GaAs & AlAs & InAs\\
\hline $m_c\,(m_0)$ & 0.0665\tablenotemark[3] &
0.150\tablenotemark[3] & 0.02226\tablenotemark[1]\\
$\gamma_1^L$ & 7.10\tablenotemark[4] & 3.76\tablenotemark[4] &
19.67\tablenotemark[3]\\
$\gamma_2^L$ & 2.02\tablenotemark[4] & 0.90\tablenotemark[4] &
8.37\tablenotemark[3]\\
$\gamma_3^L$ & 2.91\tablenotemark[4] & 1.42\tablenotemark[4] &
9.29\tablenotemark[3]\\
$E_p\,(\textrm{eV})$ & 28.0\tablenotemark[2] &
21.1\tablenotemark[3] &
22.2\tablenotemark[3]\\
$E_g\,(\textrm{eV})$ & 1.519\tablenotemark[3] &
3.130\tablenotemark[3] &
0.418\tablenotemark[3]\\
$\Delta\,(\textrm{eV})$ & 0.341\tablenotemark[3] &
0.275\tablenotemark[3] &
0.380\tablenotemark[3]\\
$E_v\,(\textrm{eV})$ & 0 & $-$0.532\tablenotemark[4] &
0.186\tablenotemark[5]\\
$\alpha$ & $-$2.27 & 0.11 & 0.24\\
$\gamma_1$ & 0.96 & 1.51 & 1.97\\
$\gamma$ & $-$0.52 & 0.09 & 0.07\\
$\chi$ & $-$1.52 & $-$0.69 & $-$0.87
\end{tabular}
\tablenotetext[1]{Ref.~\onlinecite{tab1:1}}
\tablenotetext[2]{Ref.~\onlinecite{tab1:2}}
\tablenotetext[3]{Ref.~\onlinecite{tab1:3}}
\tablenotetext[4]{Ref.~\onlinecite{tab1:4}}
\tablenotetext[5]{Ref.~\onlinecite{Stier}}
\end{table}

\newpage
\begin{table}
\caption{The 8-band effective-mass parameters of some IV-VI materials.
The VB offset $E_v$ is chosen to be zero in HgS. The parameters of
the spherical model $\alpha$, $\gamma_1$, $\gamma$, and $\chi$ for
CdS and $\chi$ for HgS are calculated from the listed here
effective-mass parameters. The parameters $m_c$, $\gamma_1^L$, and
$\gamma^L$ are not presented for HgS, because in a semimetal the
band structure is inverted and these parameters do not have their
original sense.}\label{HgS}
\begin{tabular}{lcc}
Parameters & HgS & CdS\\
\hline $m_c\,(m_0)$ & -- & 0.18\tablenotemark[6]\\
$\gamma_1^L$ & -- & 1.71\tablenotemark[6]\\
$\gamma^L$ & -- & 0.62\tablenotemark[6]\\
$E_p\,(\textrm{eV})$ & 13.2\tablenotemark[1] & 21.0\tablenotemark[5]\\
$E_g\,(\textrm{eV})$ & $-$0.190\tablenotemark[1] & 2.56\tablenotemark[3]\\
$\Delta\,(\textrm{eV})$ & 0.07\tablenotemark[3] & 0.07\tablenotemark[3]\\
$E_v\,(\textrm{eV})$ & 0 & $-$0.93\tablenotemark[4]\\
$\alpha$ & $-$1.0\tablenotemark[3] & $-$2.57\\
$\gamma_1$ & 0.35\tablenotemark[2] & $-$1.02\\
$\gamma$ & $-$0.67\tablenotemark[2] & $-$0.75\\
$\chi$ & $-$1.57 & $-$1.24
\end{tabular}
\tablenotetext[1]{Ref.~\onlinecite{Exp.HgS}}
\tablenotetext[2]{Ref.~\onlinecite{gamma.HgS}}
\tablenotetext[3]{Ref.~\onlinecite{tab2:3}}
\tablenotetext[4]{Ref.~\onlinecite{tab2:4}}
\tablenotetext[5]{Ref.~\onlinecite{tab2:5}}
\tablenotetext[6]{Ref.~\onlinecite{mass.CdS}}
\end{table}

\newpage
\begin{table}
\caption{The spin-orbital splitting of electron energy levels
($c_{\xi}=-1$); the energy-dependent electron effective masses
($c_{\xi}=-1$); the energy shift of electron and hole levels due to
finite values of $\chi$ (at $c_{\xi}=0$); and the difference of $\chi^L$
in adjacent materials for the 6$\times$6 model (at $c_{\xi}=1$). If
not indicated explicitly, $\chi\ne 0$. $E_e$ and $E_h$ are the
electron and hole ground state energies corresponding to the
states $1S_{1/2}^{(e)}$ and $1S_{3/2}^{(h)}$. The indices
``$1$'' and ``$2$'' denote the interior and exterior materials,
correspondingly.}\label{table}
\begin{tabular}{lccc}
& HgS/CdS & InAs/GaAs & GaAs/AlAs\\
& $a = 2$ nm & $a = 4$ nm & $a = 3$ nm\\
\hline
$E_{1P_{3/2}^{(e)}}-E_{1P_{1/2}^{(e)}}\,(\textrm{meV})$\tablenotemark[1]
& 2.6 (2.0) & 3.0 (1.4) & 2.7 (3.0)\\
$E_{1D_{5/2}^{(e)}}-E_{1D_{3/2}^{(e)}}\,(\textrm{meV})$\tablenotemark[1]
& 2.0 (2.3) & 3.8 (2.2) & 4.5 (7.1)\\
\hline $m_{c,1}(E_e)\,(m_0)$\tablenotemark[2] & 0.056 (\,\,--\,\,)
& 0.038 (0.022) & 0.083 (0.067)\\
$m_{c,2}(E_e)\,(m_0)$\tablenotemark[2] & 0.097 (0.180)
& 0.040 (0.067) & 0.115 (0.150)\\
\hline
$E_e-\left.E_e\right|_{\chi=0}\,(\textrm{meV})$\tablenotemark[3]
& 16.1 (6.3) & $-$3.5 ($-$3.1) & 4.8 (7.0)\\
$E_h-\left.E_h\right|_{\chi=0}\,(\textrm{meV})$\tablenotemark[3]
& 9.3 (6.3) & $-$4.1 ($-$3.1) & 4.1 (7.0)\\
\hline $\chi_1^L(E_h)-\chi_2^L(E_h)$\tablenotemark[4] & 36.28
($-$0.33) & 4.86 (0.65) & 0.73 ($-$0.83)
\end{tabular}
\tablenotetext[1]{The theoretical estimate based on the 2$\times$2
energy-dependent Hamiltonian for an electron (see Appendix~B) is
given in parentheses.} \tablenotetext[2]{The corresponding bulk
effective mass (see Tables \ref{GaAs} and \ref{HgS}) is given in
parentheses.} \tablenotetext[3]{The result of the empirical
estimate $E_{e(h)}-\left.E_{e(h)}\right|_{\chi=0}=
-\displaystyle\frac{\hbar^2}{m_0 a^2}(\chi_1-\chi_2)$ is given in
parentheses.} \tablenotetext[4]{The difference $\chi_1-\chi_2$ for
the 8-band model (see Tables \ref{GaAs} and \ref{HgS}) is given in
parentheses.}
\end{table}

% end TABLES

% begin FIGURES

\newpage

\newpage

\newpage

\newpage

\newpage

\newpage

\newpage

\newpage

\newpage

% end FIGURES


\begin{thebibliography}{49}
\bibitem{mass.CdS} Parameters $m_c$, $\gamma_1^L$, and $\gamma^L$
are computed applying the quasicubic model\cite{QC} (i.e.
$1/m_c=(2/m_{c\perp}+1/m_{c\parallel})/3$,
$\gamma_1^L=(2/m^A_{\perp}+1/m^A_{\parallel})/3$,
$\gamma^L=(1/m^A_{\perp}-1/m^A_{\parallel})/3$) to the effective
masses of wurtzite CdS. $m_{c\perp}$, $m_{c\parallel}$,
$m^A_{\perp}$, $m^A_{\parallel}$ were determined in
Ref.~\onlinecite{EM}, using the empirical pseudopotential method.
\end{thebibliography}
\end{document}